\documentclass[journal,twoside,web]{ieeecolor}

\usepackage{lcsys}
\usepackage{cite}
\usepackage{amsmath,amssymb,amsfonts}
\usepackage{graphicx}
\usepackage{textcomp}
\def\BibTeX{{\rm B\kern-.05em{\sc i\kern-.025em b}\kern-.08em
    T\kern-.1667em\lower.7ex\hbox{E}\kern-.125emX}}
\usepackage{epsfig}
\usepackage{epstopdf}

\usepackage{caption}
\usepackage{subcaption}
\usepackage{float}

\usepackage[utf8]{inputenc} 
\usepackage[T1]{fontenc}    
\usepackage{hyperref}       
\usepackage{url}            
\usepackage{booktabs}       
\usepackage{nicefrac}       
\usepackage{microtype}      
\usepackage{lipsum}
\usepackage{graphicx}
\graphicspath{ {./images/} }

\usepackage{cite}
\usepackage{amsmath,amssymb,amsfonts,mathrsfs}
\usepackage{epstopdf}
\usepackage{caption}
\usepackage[font=scriptsize]{caption}

\newcommand*{\Scale}[2][4]{\scalebox{#1}{$#2$}}%

\newtheorem{thm}{Theorem}
\newtheorem{lem}{Lemma}
\newtheorem{prop}{Proposition}
\newtheorem{cor}{Corollary}
\newtheorem{defn}{Definition}

\newtheorem{rem}{Remark}

\newtheorem{assm}{Assumption}

\usepackage{textcomp}
\usepackage{lipsum}                     
\usepackage{xargs}                      
\begin{document}

\title{A Convex Optimization Approach for Control of Linear Quadratic Systems with Multiplicative Noise via System Level Synthesis}

\author{Majid Mazouchi and Farzaneh Tatari and Hamidreza Modares, \IEEEmembership{Senior Member, IEEE}
\thanks{}
\thanks{ 
Majid Mazouchi and Farzaneh Tatari and Hamidreza Modares are with the Mechanical Engineering Department, 
  Michigan State University, 
  East Lansing, MI, 48863 (e-mails: mazouchi@msu.edu, modaresh@msu.edu). }
}

\maketitle

\begin{abstract}
This paper presents a convex optimization-based solution to the design of state-feedback controllers for solving the linear quadratic regulator (LQR) problem of uncertain discrete-time systems with
multiplicative noise. To synthesize a tractable solution, the recently developed system level synthesis (SLS) framework is leveraged. It is shown that SLS shifts the controller synthesis task from the design of a robust controller to the design of the entire set-valued closed-loop system responses. To this end, the closed-loop system response is entirely characterized by probabilistic set-valued maps from the additive noise to control actions and states. A bi-level convex optimization over the achievable set-valued closed-loop responses is then developed to optimize the expected value of the LQR cost against the worst-case closed-loop system response. The solution to this robust optimization problem may be too conservative since it aims at enforcing the design constraints for all possible system realizations. To deal with this issue, the presented optimization problem is next reformulated as a chance-constrained program (CCP) in which the guarantees are not intended in a deterministic sense of satisfaction against all possible closed-loop system responses, but are instead intended in a probabilistic sense of satisfaction against all but a small fraction of the system responses. To approximately solve the CCP without the requirement of knowing the probabilistic description of the uncertainty in the system matrices, the so-called scenario optimization approach is employed,  which provides probabilistic guarantees based on a finite number of system realizations and results in a convex optimization program with moderate computational complexity. Finally, numerical simulations are presented to illustrate the theoretical findings.
\end{abstract}



\section{Introduction}
\label{sec:introduction}
\IEEEPARstart{L}{inear} 
 Quadratic Regulator (LQR) problem has been one of the most mature and popular methods for control of linear systems \cite{kalman1960contributions,anderson1990optimal}, and has been widely leveraged in a variety of applications, such as aerospace, robotics, finance, etc. The aim of the LQR problem is to design a state-feedback controller that minimizes a convex quadratic cost related to the control of a linear dynamical system. Even though the LQR problem for linear systems with known dynamics is a mature control technology, the presence of system uncertainties brings about challenges of robustness, efficiency, and feasibility of control solutions. The robust LQR problem for systems with additive noise has been investigated in an anthology of papers, see \cite{doyle1978guaranteed,scampicchio2021stable,anderson2019system,xue2021data}, and references therein.  The min-max model predictive control
(MPC) framework  \cite{bemporad1999robust,raimondo2009min,saltik2018outlook} also provides approximate robust solutions for constrained LQR problems. Data-based LQR has also been recently considered in \cite{LQRdata1,LQRdata2,LQRdata3}, and the robustness of the solutions is either accounted for during the design or analyzed a posteriori. 

Most results on the robust LQR problem have been developed for systems with additive noise, and the LQR control design for systems with multiplicative noise is scarce \cite{gravell2019learning,coppens2020data,gravell2020robust,pang2022robust}. However, it is of great importance to design robust LQR controllers for systems with multiplicative noise to account for dynamics perturbations of the systems. Systems with multiplicative noise explicitly incorporate inherent stochasticity in the system dynamics. Moreover, as shown in this paper, data-based control of unknown systems with additive noise can also amount to control of identified systems that are characterized by systems with multiplicative noise. Robust LQR control of systems with multiplicative noise is considered in \cite{jongeneel2019robust}. Policy gradient is leveraged to learn the LQR solution, which leads to non-convex optimization problems. Moreover, the input and the dynamics noises are assumed to be mutually independent. However, as shown in this paper, these noises can be dependent on some applications. Moreover, robust mean-square stability is provided in \cite{jongeneel2019robust}. Robust LQR control approaches for a system with multiplicative noise require optimizing the performance function for the worst-case system realization. This, however, can be overly conservative, especially if the support of the uncertainty is large.

The recently introduced system level synthesis (SLS) framework \cite{anderson2019system,xue2021data,dean2020sample} provides powerful tools to transform the design of a robust LQR controller into a convex optimization problem. The core concept behind the SLS framework scheme is that it transforms the control design over the linear feedback control gains to closed-loop system responses and provides an explicit link between them. The robust form of the SLS framework \cite{chen2019system,lian2021system,xue2021data,chen2020robust,matni2020robust} allows for an explicit mapping from the model uncertainty to the system behavior, providing an explicit characterization of the joint effects of additive disturbances and model errors when solving robust LQR problem \cite{gravell2019learning,wonham1967optimal}. The rationale behind the SLS framework is both practical and applicable to many settings, and since its beginning, there have been many extensions developed, such as works on MPC \cite{bujarbaruah2020robust,li2021frontiers,sieber2021system,chen2020robust,chen2021system}, dynamic programming \cite{tseng2020system}, data-driven adaptive control \cite{xue2021data,lian2021system}, and so forth. However, the SLS framework for solving the LQR problem for systems with multiplicative noise has not  been considered in the literature, despite its importance.

To sidestep the issues of conservativeness of robust solutions and their possible infeasibility, one alternative approach is to interpret robustness in a probabilistic sense, in which the guarantees of constraints fulfillment are
intended in the probabilistic sense (satisfying most uncertain instances) rather than the deterministic sense (satisfaction against all possible uncertain outcomes). That is, constraint violation is allowed with a low probability. This leads to
stochastic optimization problems, which are typically called
chance-constrained programming (CCP) problems \cite{prekopa2013stochastic,calafiore2011research,geng2019data,van2015distributionally,coulson2021distributionally}. CCP has been widely used in a wide range of disciplines, e.g., in finance \cite{stojanovic2003stochastic}, control \cite{schildbach2013linear}, and so forth. With the
exception of a few special cases  \cite{Lemarchal2006SBL}, however, CCP problems are computationally intractable (i.e., NP-hard) since they require the computation of multi-dimensional probability integrals \cite{NemirovskiArkadi2006ConvexAO}.
Scenario approach \cite{campi2009scenario,Camp2021TheSA,Nemirovski2006ScenarioAO,Calafiore2006TheSA} is a simple yet promising method for approximately solving chance-constrained optimization. To this aim, the scenario approach employs a dataset with some samples (so-called scenarios) from the set of uncertain parameters and requires the constraints to be satisfied for each scenario. A prominent feature of the scenario approach is its generality and tractability, as well as the fact that it requires no assumptions apart from constraint convexity. 

This paper presents a convex optimization approach to design state-feedback controllers for solving the linear quadratic regulator (LQR) problem of uncertain discrete-time systems with
multiplicative noises. To synthesize a tractable solution, the recently developed SLS framework is adopted to shift the controller synthesis task from the design of a robust controller to the design of the entire set-valued closed-loop system responses with high probability. To this end, the closed-loop system responses are entirely characterized by probabilistic set-valued maps from the additive noise to control actions and states. It is then shown that the robust SLS formulation for LQR control design for systems with multiplicative noise amounts to a bi-level program. The solution to this bi-level program, however, may be computationally expensive and also too conservative, since it aims at enforcing the design constraints for all possible system realizations. To sidestep this issue, the presented optimization problem is next reformulated as a convex chance-constrained program (CCP) in which the guarantees are not intended in a deterministic sense of satisfaction against all possible uncertainty outcomes, but are instead intended in a probabilistic sense of satisfaction against all but a small fraction of the systems. The scenario optimization approach is then employed to provide probabilistic guarantees based on a finite number of system realizations. The resulting optimization is a convex optimization program with moderate computational complexity. A simulation example is provided to show the effectiveness of the presented approach. \vspace{6pt}

{\bf{Notations.}} The following notations will be used throughout this paper.
$I_{n}$ denotes the $n \times n$ identity matrix.
$0_{m \times n}$ denotes the matrix of all zero entities.
Note that, sometimes, the subscripts $_n$ and $_{m \times n}$ will be omitted for notational brevity if there is no confusion.
$\mathbb{R}$ and $\mathbb{N}$ represent the sets of real numbers and natural numbers, respectively. The $n$-dimensional Euclidean space is denoted by $\mathbb{R}^{n}$. 
$\otimes$ denotes the Kronecker product.
$H^\top$ denotes transpose of a matrix $H$. 
The notation $diag( . )$ refers to a diagonal matrix with the argument(s) on the diagonal.
 $blkdiag \left(A_{1}, A_{2}, \ldots, A_{k}\right)$ represents the block diagonal matrix with diagonal blocks $A_{1}, A_{2}, \ldots, A_{k}$.
$\left[ {\rm E} \right]_i^{r}$ and $\left[ {\rm E} \right]_j^{c}$ denote the $i$-th row and $j$-th column of ${\rm E}$, respectively, and ${[E]_{ij}^{cr}}$ denotes the block-entry on $i$-th row and $j$-th column of $E$.
If $E$ and $D$ are matrices (or vectors) of the same dimensions, then $E(\leq, \geq) D$ implies componentwise inequality, i.e., $[E]_{ij}^{cr}(\leq, \geq) D_{ij}^{cr}$ for all $i$ and $j$.
We assume that all uncertainty entering matrices are random variables defined on a probability space $(\Delta, \mathcal{D}, \mathbb{P})$, with $\Delta$ the sample space, $\mathcal{D}$ its associated $\sigma$-algebra and $\mathbb{P}$ the probability measure. When a random variable $X$ is distributed normally with mean $v$ and variance $\sigma^{2}$, we use the notation $X \sim \mathscr{N}\left(v, \sigma^{2}\right)$. $\mathbb{E}[X]$ and $\Sigma_{X}=\mathbb{E}\left[(X-\mathbb{E}[X])(X-\mathbb{E}[X])^\top \right]$ denotes, respectively, the expectation and the covariance of $X$.
Let $\vec x: \Delta \rightarrow \mathbb{R}^{n}$ be  a random vector defined on $(\Delta, \mathcal{D}, \mathbb{P}) .$ With some abuse of notation, we will write $\vec x \in \mathbb{R}^{n}$ to state the dimension of this random vector. Let $\mathbb{P}_{\vec x}$ denote the distribution of $\vec x$, i.e., $\mathbb{P}_{\vec x}(A)=\mathbb{P}[\vec x \in A]$. Then, a trajectory $\left\{\vec x_{i}\right\}_{i=1}^{N}$ of identically and independently distributed (i.i.d.) copies of $\vec x$ is defined by the distribution it induces. That is, for any $A_{0}, \ldots, A_{N} \in \mathcal{D}$, we define $\mathbb{P}_{\vec x_{0}, \ldots, \vec x_{N}}\left(A_{0} \times \cdots \times\right.$ $\left.A_{N}\right):=\mathbb{P}\left[\vec x_{0} \in A_{0} \wedge \cdots \wedge \vec x_{N} \in A_{N}\right]=\prod_{i=0}^{N} \mathbb{P}_{\vec x}\left(A_{i}\right)$. This definition can be extended to infinite trajectories $\left\{\vec x_{i}\right\}_{i \in \mathbb{N}}$ by Kolmogorov's existence theorem \cite{Krmer1995ProbabilityM}.
${x}_{[k, m]}$ is used as shorthand for the signal $\left[x^{\top}(k) x^{\top}(k+1) \cdots x^{\top}(m)\right]^{\top}$. A linear, causal operator $\mathbf{R}$ defined over a horizon of $T$ is represented by
$$\mathbf{R}=\left[\begin{array}{cccc}R^{0,0} & & & \\ R^{1,1} & R^{1,0} & & \\ \vdots & \ddots & \ddots & \\ R^{T, T} & \cdots & R^{T, 1} & R^{T, 0}\end{array}\right],$$
where $R^{i, j} \in \mathbb{R}^{p \times q}$ is a matrix of compatible dimensions. The set of such matrices are denoted by $\mathcal{L}^{T, p \times q}$
 and the superscript $p \times q$ is dropped when it is clear from the context. 
An operator $\mathbf{R} \in \mathcal{L}^{T}$ acts on a signal $\sigma_{[0, T-1]}$ through multiplication, i.e., $x_{[0, T-1]}=\mathbf{R} \sigma_{[0, T-1]}$.
$\text{vec}\{A\}= [a_{11}\,\, a_{21}, \,\, \dots \,\, a_{n1} \,\, a_{21} \,\, a_{22} \,\, \dots \,\, a_{nm}]^{\top}$ 
is the vectorization of the matrix $A \in \mathbb{R}^{n \times m}$.
Finally, $\|\cdot\|_{F}$, $\|.\|$, $\|\cdot\|_{V}$  denote the Frobenius norm, the Euclidean norm, and the weighted Euclidean norm (i.e, $\|x_t\|^2_{V}=x_t^{\top} V x_t$), respectively.

\section{preliminaries and problem formulation}
\label{sec:2}
\subsection{System Level Synthesis}
Consider the following discrete-time linear time invariant (LTI) systems of the form
\begin{align} 
\Scale[1]{{x_{t + 1}} = {A}{x_t} + {B}{u_t} + {\mathcal{W}_t} } \quad \text{for} \quad  t=0,1, \ldots, T    \label{LTI}
\end{align}
where ${x_t} \in \mathbb{R}{^n}$, ${u_t} \in \mathbb{R}{^m}$, and $\mathcal{W}_{t} \in \mathbb{R}{^n} $ is an exogenous disturbance or process noise. It is assumed here that the pair $(A,B)$ is controllable.

Letting the control input $u_t$ to be a causal linear time-varying state-feedback controller, i.e., $u_t = K_t{{x}_{[0,t]}}$ for some linear maps $K_t$, the SLS casts the controller synthesis task as an optimization over the entire realizable closed-loop behaviors of the system over the horizon $t=0, \ldots, T-1$. To see this, consider the finite-horizon LQR problem formulated as
\begin{align}
\Scale[0.9]{  \begin{array}{*{20}{l}}
{ \mathop {{\rm{min}}}\limits_{{u_t}} \quad \mathbb{E} \left[ {\sum\limits_{t = 0}^{T-1 } {x_t^\top} Q{x_t} + u_{t}^\top R{u_{t}} } \right]},\\
{{\rm{s}}.{\rm{t}}.\,\left\{ \begin{array}{l}
{x_{t + 1}} = A{x_t} + B{u_t} + {{\cal W}_t},\\
{u_t} =  K_t{{x}_{[0,t]}}
\end{array} \right.}
\end{array}} 
\label{eq:lqr}
\end{align}
where $Q\succeq 0$ and $R \succ 0$. Utilizing ${u_t} = K_t{{x}_{[0,t]}}$, the closed-loop dynamics can be compactly (expressed as block matrix operations) written as \cite{anderson2019system} 
\begin{align}
\Scale[1]{{x_{[0,T-1]}} = {\cal Z}({\cal A} + {\cal B}{\bf{K}}){x_{[0,T-1]}} + \mathbf{w} },
\label{eq:5*}
\end{align}
where
$\mathbf{w}=\left[\begin{array}{c}x_{0} , {{\cal W}}_0 ,{{\cal W}}_1 , \hdots , {{\cal W}}_{T-2}\end{array}\right]^\top$,
%
 $\mathcal{A}:= blkdiag(A,...,A,0)$, and $\mathcal{B}:= blkdiag(B,...,B,0) $,  $\mathcal{Z}$ is the block-downshift operator, i.e., a matrix with identity matrices along the first block subdiagonal and zeros elsewhere, and ${\bf{K}} \in \mathcal{L}^{T, m \times n}$ represents the block matrix operator for the causal linear time-varying state-feedback controller. Rewriting \eqref{eq:5*}, the closed-loop map (i.e., system behavior) from disturbance to state and control input, respectively, is given by \cite{anderson2019system} 
\begin{align}
\Scale[1]{ \begin{aligned}
{\bf x}:= x_{[0, T-1]} &=(I-\mathcal{Z}(\mathcal{A}+\mathcal{B} {\bf{K}}))^{-1} \mathbf{w}={\bf \Phi}_{x} \mathbf{w}, \\
{\bf u}:= u_{[0, T-1]} &={\bf{K}}(I-\mathcal{Z}(\mathcal{A}+\mathcal{B} {\bf{K}}))^{-1} \mathbf{w}={\bf \Phi}_{u} \mathbf{w},
\end{aligned}}
\label{eq:6*}
\end{align}
where
$\left\{{\bf \Phi}_{x}, {\bf \Phi}_{u}\right\}$ are two block-lower-triangular matrices called as the closed-loop system response and one realization of the controller is given by ${\bf{K}} ={ \bf \Phi}_{u} {\bf \Phi}_{x}^{-1} \in$ $\mathcal{L}^{T}$.

The following proposition will prove useful in the rest of the development.
\begin{prop}\label{prop:1} \cite{anderson2019system}
For a system \eqref{LTI} with state-feedback control law ${\bf{K}} \in \mathcal{L}^{T}$, i.e., $u_{[0, T-1]}= {\bf{K}} x_{[0, T-1]}$, the following statements are true
\begin{enumerate}
\item [1:] The affine subspace defined by
\begin{align}
\Scale[1]{\left[\begin{array}{ll}(I-\mathcal{Z} \mathcal{A}) & -\mathcal{Z} \mathcal{B}\end{array}\right]\left[\begin{array}{c}{\bf \Phi}_{x} \\ {\bf \Phi}_{u}\end{array}\right]=I,}
\label{eq:8*}
\end{align}
with ${\bf \Phi}_{x} \in \mathcal{L}^{T, n \times n}$ and ${\bf \Phi}_{u} \in \mathcal{L}^{T, m \times n}$, parameterizes all possible system responses from $\mathbf{w} \rightarrow\left(\mathbf{x}, \mathbf{u}\right)$; \vspace{3pt}
\item [2:] For any causal linear operators ${\bf \Phi}_{x}, {\bf \Phi}_{u}$ satisfying \eqref{eq:8*}, the controller ${\bf{K}}={\bf \Phi}_{u} {\bf \Phi}_{x}^{-1}$ is internally stabilizing and achieves the desired closed-loop responses \eqref{eq:6*}.
\end{enumerate}
\end{prop}

Using Proposition \ref{prop:1}, the LQR problem \eqref{eq:lqr} can be reformulated as a convex optimization problem in terms of the system responses $\left\{{\bf \Phi}_{x}, {\bf \Phi}_{u}\right\}$ as  \cite{anderson2019system}
\begin{align} 
\Scale[1]{  \begin{array}{l}
\mathop {{\mathop{\rm min}\nolimits} }\limits_{{{{\bf \Phi }}_x},{{{\bf \Phi }}_u}} {\left\| {\left[ {\begin{array}{*{20}{l}}
{{{\cal Q}^{1/2}}}&{}\\
{}&{{{\cal R}^{1/2}}}
\end{array}} \right]\left[ {\begin{array}{*{20}{l}}
{{{{\bf \Phi }}_x}}\\
{{{{\bf \Phi }}_u}}
\end{array}} \right]{\bf w}} \right\|_F^2}\\
\,\,\,\,\,s.t.\,\,\,\left[ {\begin{array}{*{20}{c}}
{\left( {I - {\cal Z}{\cal A}} \right)}&{ - {\cal Z}{\cal B}}
\end{array}\,\,\,\,} \right]\left[ {\begin{array}{*{20}{l}}
{{{{\bf \Phi }}_x}}\\
{{{{\bf \Phi }}_u}}
\end{array}} \right] = I
\end{array} }
\label{eq:9*}
\end{align}
where $\mathcal{Q}:=I_{T} \otimes Q$ and $ \mathcal{R}:=I_{T} \otimes R$. 

The constraint in \eqref{eq:9*} encores the closed-loop system responses to obey the system dynamics and the cost function is optimized over all achievable closed-loop responses $\{{{{{\bf \Phi }}_x}},{{{{\bf \Phi }}_u}}\}$.  The control gain is then given by ${\bf{K}} ={\bf \Phi}_{u} {\bf \Phi}_{x}^{-1}$. 

\subsection{Motivation and Problem Formulation}
While SLS has significant benefits in terms of scalability and tractability, its application is typically limited to systems with additive noise. However, systems with multiplicative noise are common in the real world. Moreover, as shown later, even systems with additive noise and unknown dynamics amounts to systems with multiplicative noise after being identified through collected data. 
%

Consider linear systems in the form of 
\begin{align} 
    x_{t+1}=A(\delta_t) x_t+ B(\delta_t) u(t)+{{\cal W}_t},     \label{eq:13*}
\end{align}
where $A(\delta_t)$ and $B(\delta_t)$ are the uncertain system and input dynamics at time $t$, respectively, with $\delta_t$ is a random variable. Moreover, ${x_t} \in \mathbb{R}{^n}$, ${u_t} \in \mathbb{R}{^m}$, and $\mathcal{W}_{t} \in \mathbb{R}{^n} $ is an exogenous random process.
In many applications, the time-varying dynamics $A(\delta_t)$ and $B(\delta_t)$ can be represented in the following  form
\begin{align}
 &   A(\delta_t) := {A_0} + \sum\limits_{i = 1}^{n_{\delta}} {\delta _{}^{(i)}} {A_i},  \label{mAB} \\
    & B(\delta_t) := {B_0} + \sum\limits_{i = 1}^{n_{\delta}} {\delta _{}^{(i)}} {B_i}, \label{mABB}
\end{align}
where $A_0$ and $B_0$ are the nominal dynamics and the summations capture multiplicative noise terms in which $A_i, \,\, B_i \,\, i=1,...,n_{\delta}$ are known and the scalar random variables $\delta^{(i)}$ are modeled by the i.i.d. zero-mean, mutually independent random noises with the variance $\sigma_i^2$.

This representation explicitly incorporates the model stochasticity inherited in many real-world applications. Nevertheless, as shown next, even for systems with only additive noise that do not have any inherent stochasticity in their dynamics, i.e., $A(\delta_t)=A $ and $ B(\delta_t)=B$, $\forall t$, the learned dynamics for unknown matrices $A$ and $B$ through collected data will be in the form of systems with multiplicative noise. Therefore, the controller must be designed for a system in the multiplicative form even though the original system is in the form of systems with additive noise.  \vspace{6pt}

\begin{lem} Consider the LTI system \eqref{LTI} (i.e., the system  \eqref{eq:13*} with $ A(\delta_t)=A $ and $  B(\delta_t)=B$, $\forall t$ and let the additive noise ${{\cal W}_t}$ be a zero-mean Gaussian noise with the variance
$diag(\alpha_1,...,\alpha_n)$ where $\alpha_1, ..., \alpha_n \in \mathbb{R}$.
Let the state-input data collected from the system be $x_{[0,T]}$ and $u_{[0,T-1]}$. Let 
\begin{align} 
    \left[ \begin{array}{l}
{u_{[0,T-1]}}\\
{x_{[0,T-1]}}
\end{array} \right]      \label{dataR}
\end{align}
be full-row rank. Then, the identified system is in the form \eqref{eq:13*} of systems with $\hat A(\delta_t) $ and $ \hat B(\delta_t)$ and some multiplicative noise $w_i=\mathcal{N}(0,\alpha_i), \, i=1,..., n$. \vspace{6pt}

\noindent \textbf{Proof.} Let $\hat A$ and $\hat B$ be the estimated dynamics for $A$ and $B$, respectively. Based on the system \eqref{LTI}, and using the collected data, one has
\begin{align} 
& x_{[1,T]} = [\hat B{\rm{  }}\,\,\,\,{{\hat{A}}}]\left[ \begin{array}{l}
{u_{[0,T-1]}}\\
{x_{[0,T-1]}}
\end{array} \right]+{{\cal W}_{[0,T-1]}},  \label{sysdata}
\end{align}
If the matrix \eqref{dataR} is full-row rank, then there exists a right inverse $[V_1 \,\,\, V_2]$ such that 
\begin{align} 
\left[ \begin{array}{l}
{u_{[0,T-1]}}\\
{x_{[0,T-1]}}
\end{array} \right]
 [V_1 \,\,\, V_2]=
I,  \label{inv}
\end{align}
Multiplying the both sides of \eqref{sysdata} by $[V_1 \,\,\, V_2]$, one has
\begin{align}
 x_{[1,T]}[V_1 \,\,\, V_2] = [\hat B{\rm{  }}\,\,\,\,{{\hat{A}}}]+{{\cal W}_{[0,T-1]}}[V_1 \,\,\, V_2], \label{eq1}
\end{align}
Therefore, 
\begin{align} 
  & \nonumber \hat A=x_{[1,T]} V_1-{{\cal W}_{[0,T-1]}} V_1=x_{[1,T]} V_1+\sum_{i=1}^n w_{i} {\hat A}_i,
\end{align}
which is in the form of \eqref{mAB} with $A_0=x_{[1,T]} V_1$,
 and ${\delta _{}^{(i)}}=w_i=\mathcal{N}(0,\alpha_i)$,
where $w_i$ is the $i$-th component of ${\cal W}$, and $ {\hat A}_i \in \mathbb{R}^{n \times n}$ is given by
\begin{align} 
{\hat A}_i=\begin{bmatrix}
 0 & 0 & \dots & 0 \\
\vdots & \vdots & \vdots & \vdots \\
\sum(V_1(:,1))&\sum(V_1(:,2))&\dots&\sum(V_1(:,n))\\
\vdots & \vdots & \vdots & \vdots\\
0 & 0 & \dots & 0
\end{bmatrix}, \label{eq:13}
\end{align}
where $\sum(V_1(:,i))$ returns the column sum of the $i$ column of the matrix $V_1$. That is, $ {\hat A}_i$ is an $n \times n$ matrix for which only its $i$ row is nonzero and is formed by the column sums of the columns of $V_1$. Similarly, for the input dynamics, one has
\begin{align} 
  \hat B=x_{[1,T]} V_2-{{\cal W}_{[0,T-1]}} V_2=x_{[1,T]} V_2+\sum_{i=1}^n w_i {\hat B}_i,
\end{align}
where ${\hat B}_i$ is defined similarly to \eqref{eq:13}, with $V_1$ being replaced by $V_2$.
\end{lem} \hfill $\square$

\begin{rem}
Note that a bulk of research has been performed to construct reasonable (not necessarily optimal, however) confidence intervals for learned $\hat A$ and $\hat B$ in the form of $\|\hat A-A\| \le \epsilon_A(N,\bar \delta)$ and $\|\hat B-B\| \le \epsilon_B(N,\bar \delta)$ in terms of the number of samples $N$ and the confidence level $\bar \delta$ \cite{systemID1,systemID2,systemID3,systemID4,systemID5,systemID6,systemID7,systemID8,systemID9,systemID10,systemID11}. Robust LQR scheme  can then be  leveraged to control the systems  with additive noise and uncertain dynamics. However, quantifying the uncertainty is challenging as it depends on the system properties that are not known (e.g., finite time controllability Gramians for the control and noise inputs), and also it requires discarding all the trajectory data except the very last state transition to exploit the independence across trajectories \cite{LQRdata3}. By contrast, the multiplicative modeling of the identified model can completely characterize the uncertainly using only collected data used for learning. 
\end{rem}

Now consider the system with multiplicative noise \eqref{eq:13*}-\eqref{mABB}. The following problem is considered and sample-based efficient convex optimization-based solutions with high-probability guarantees and less-conservativeness (compared to robust approaches) will be presented using SLS in this paper. 
\begin{align}
\Scale[0.9]{ \text{P} \, 1: \begin{array}{*{20}{l}}
{ \mathop {{\rm{min}}}\limits_{{u_t}} \quad \mathbb{E} \left[ {\sum\limits_{t = 0}^{T-1 } {x_t^\top} Q{x_t} + u_{t}^\top R{u_{t }} } \right]}\\
{{\rm{s}}.{\rm{t}}.\,\left\{ \begin{array}{l}
{x_{t + 1}} = ({A_0} + \sum\limits_{i = 1}^{n_{\delta}} {\delta _{}^{(i)}} {A_i}){x_t} + ({B_0} + \sum\limits_{i = 1}^{n_{\delta}} {\delta _{}^{(i)}} {B_i}) + {{\cal W}_t}\\
{u_t} = K_t{{x}_{[0,t]}}
\end{array} \right.}
\end{array}} 
\label{eq:lqrm}
\end{align}
where the cost function is similar to the one defined in \eqref{eq:lqr}. \vspace{6pt}

\begin{rem} A policy gradient solution is presented for solving problem P2 in \cite{gravell2019learning}. Even though the results are elegant, non-convex optimization problems must be solved. Moreover, robust mean-square stability guarantees are provided (as defined below), which can be overly conservative. Finally, the input and the dynamics noises are assumed to be mutually independent in \cite{gravell2019learning}. However, as it was shown in the motivation section, these noises can be dependent in some applications. 
\end{rem} \vspace{6pt}

\begin{defn} (Mean-square stability) \cite{KOZIN196995}. The system in \eqref{eq:13*} is mean-square stable if and only if $\text{lim}_{t \rightarrow \infty} \mathbb{E} \{x_t \, x_t^{\top} \}=0,  \,\, \forall \|x_0\| < \infty $.
\end{defn}


\section{system level synthesis for systems with multiplicative noise: a robust control formulation}

This section presents an SLS-based convex optimization formulation for solving Problem P2. The conservativeness of the solution will be discussed, and relaxed sample-based solutions will be presented for it in the subsequent sections. \vspace{6pt}

\begin{lem} \label{setlem} Let the controller be $u_t=K x_t$. Then, with probability of at least $1-\epsilon$, the next state $x_{t+1}$ of the system \eqref{eq:13*}-\eqref{mABB} lies in the following ellipsoidal set 
\begin{align} 
     \varepsilon (V_x,1)=\Big\{x:\big(x-\bar x_{t+1}\big) \,\, V_x \,\, \big(x-\bar x_{t+1} \big)^{\top} \le 1 \Big\}, \label{set}
\end{align}
where 
\begin{align} 
    \bar x_{t+1}=(A_0+B_0 \, K) \bar x_t+\mathcal{W}_t,     \label{nom}
\end{align}
is the nominal closed-loop LTI system, and
\begin{align} 
 \nonumber   V_x=\frac{1}{{n_{\delta}}+2 \sqrt{{n_{\delta}}\,\, \text{log}{\frac{1}{\epsilon}}}+2\text{log}{\frac{1}{\epsilon}}} \times \label{V} \\
 \sum_{i=1}^{n_{\delta}}  \sigma_i^2 \,\, (A_i+B_i \, K) \,\, x_t x_t^{\top} (A_i+B_i \, K)^{-1}.
\end{align} 
Moreover, the next input $u_{t+1}$ lies in the follwing ellipsoidal set 
\begin{align} 
     \varepsilon (V_u,1)=\Big\{u:\big(u-\bar u_{t+1}\big) \,\, V_u \,\, \big(u-\bar u_{t+1} \big)^{\top} \le 1 \Big\},     \label{sety}
\end{align}
where $\bar u_{t}=K \bar x_t, $ $V_u$ is similar to $V_x$ with $x_t \, x_t^T$ being replaced by $u_t \, u_t^T$.

\noindent \textbf{Proof.} Using $u_t=K x_t$ in \eqref{eq:13*}, the random variable $\tilde x_{t+1}=x_{t+1}-\bar x_{t+1}$ has the following covariance
\begin{align}
  \nonumber  &\mathbb{E}[\tilde x_{t+1} \tilde x_{t+1}^{\top}]=\\
\nonumber   &\mathbb{E}\Big[\Big(\sum_{i=1}^{n_{\delta}} \delta_t^{(i)}  (A_i+B_i \, K) x_t \Big)  \Big(\sum_{i=1}^{n_{\delta}} \delta_t^{(i)}  (A_i+B_i \, K) x_t \Big)^{\top}\Big]=\\
& \sum_{i=1}^{n_{\delta}}  \sigma_i^2 \,\, (A_i+B_i \, K) \,\, x_t x_t^{\top} (A_i+B_i \, K).
\end{align}
Therefore, using the Chernoff inequality, with probability at least $1-\epsilon$, one has \cite{confbound}
\begin{align}
  \nonumber &\tilde x_{t+1} \Big(\sum_{i=1}^{n_{\delta}}  \sigma_i^2 \,\, (A_i+B_i \, K) \,\, x_t x_t^{\top} (A_i+B_i \, K) \Big)^{-1} \tilde x_{t+1}^{\top}  \\ 
& \quad \quad \le     {n_{\delta}}+2 \sqrt{{n_{\delta}}\,\, \text{log}{\frac{1}{\epsilon}}}+2\text{log}{\frac{1}{\epsilon}},
   \end{align}
or equivalently, with probability at least $1-\epsilon$,
\begin{align}
   &\tilde x_{t+1} V_x \tilde x_{t+1}^{\top} \le 1.
   \end{align}
The proof for the set of the next inputs is also similar. This completes the proof. \end{lem} \hfill $\square$ \vspace{6pt}

Before moving to the next theorem, which characterizes the entire set-valued closed-loop responses of systems with multiplicative noise, define
\begin{align}
 &{\bf \bar \Phi}_{x}:={\left( {{I} - { {\cal Z}}\left( {{{\cal A}_0} + {{\cal B}_0}{{\bf{K}}}} \right)} \right)^{ - 1}},
\label{phix1} \\
 &{\bf \bar \Phi}_{u}:={\bf{K}} {\left( {{I} - { {\cal Z}}\left( {{{\cal A}_0} + {{\cal B}_0}{{\bf{K}}}} \right)} \right)^{ - 1}}.
\label{phiu1}
\end{align}
where ${{\cal A}_0}= \left[ {\begin{array}{*{20}{c}}
{{I_{T - 1}}}&{}\\
{}&0
\end{array}} \right] \otimes {A_0}$ and ${{\cal B}_0}= \left[ {\begin{array}{*{20}{c}}
{{I_{T - 1}}}&{}\\
{}&0
\end{array}} \right] \otimes {B_0}$. \vspace{6pt}

\begin{thm} \label{SLS1} Consider the affine subspace defined by
\begin{align}
\Scale[1]{\left[\begin{array}{ll}(I-\mathcal{Z} \mathcal{A
}_0) & -\mathcal{Z} \mathcal{B}_0\end{array}\right]\left[\begin{array}{c}   {\bf \bar \Phi}_{x} \\ 
{\bf \bar \Phi}_{u}\end{array}\right]=I,}
\label{SLSnom}
\end{align}
with ${\bf \bar \Phi}_{x} \in \mathcal{L}^{T, n \times n}$ and ${\bf \bar \Phi}_{u} \in \mathcal{L}^{T, m \times n}$. Then, the set defined below parametrizes the entire high-confidence set-valued closed-loop responses from $\mathbf{w} \rightarrow (\mathbf{x},\mathbf{u})$ 
\begin{align} 
     \Big\{\mathbf{(x,u)}:\|\mathbf{x}- {\bf \bar \Phi}_{x} \mathbf{w}\|^2_{V_x}  \le 1, \, \|\mathbf{u}- {\bf \bar \Phi}_{u} \mathbf{w}\|^2_{V_u}  \le 1 \Big\},     \label{ineqres}
\end{align}
where  $\mathbf{x}=x_{[0, T-1]}$, $\mathbf{w}$ is defined in \eqref{eq:5*}  and  $V_x$ is defined in \eqref{V} and $V_u$ is the same as $V_x$ with with $x_t \, x_t^T$ being replaced by $u_t \, u_t^T$. \vspace{3pt}

\noindent \textbf{Proof.} 
The nominal system \eqref{nom} is an LTI system with additive noise and thus, based on \cite{anderson2019system}, the affine subspace \eqref{SLSnom} with \eqref{phix1} and \eqref{phiu1} parameterizes all possible system responses from $\mathbf{w} \rightarrow (\mathbf{\bar x},\mathbf{\bar u})$ with the transfer functions $ \{{\bar \Phi}_{x},{\bar \Phi}_{u}\}$. Therefore, using $\mathbf{ \bar x}= {\bar \Phi}_{x} \mathbf{w}$ and $\mathbf{ \bar u}= {\bar \Phi}_{u} \mathbf{w}$, and the results of Lemma~\ref{setlem} completes the proof. \end{thm} \hfill $\square$ \vspace{6pt}

\begin{thm} A robust SLS formulation of Problem P1 is given by
\begin{align} 
\Scale[0.88]{ \text{P} \, 2: \begin{array}{l}
{\mathop {{\rm{min}}}\limits_{{{{\bf \bar \Phi }}_x},{{\bf \bar {\Phi }}_u}}  \mathop {\max }\limits_
{\|{{{\bf{\tilde x}}}\|_{V_{x}} \le 1},{\|{{\bf{\tilde u}}}}\|_{V_{u}} \le 1} } {\left\| {\left[ {\begin{array}{*{20}{l}}
{{{\cal Q}^{1/2}}}&{}\\
{}&{{{\cal R}^{1/2}}}
\end{array}} \right]\left[ {\begin{array}{*{20}{l}}
{{{{\bf \bar \Phi }}_x}}\\
{{{{\bf \bar \Phi }}_u}}
\end{array}} \right]}{\bf w}+\left[ {\begin{array}{*{20}{l}}
{\bf{\tilde x}}\\
{{\bf{\tilde u}}}
\end{array}} \right] \right\|_F^2}\\
\,\,\,\,\,s.t.\,\,\,
\left[ {\begin{array}{*{20}{c}}
{\left( {I - {\cal Z}{\cal A}_0} \right)}&{ - {\cal Z}{\cal B}_0}
\end{array}\,\,\,\,} \right]\left[ {\begin{array}{*{20}{l}}
{{{{\bf \bar \Phi }}_x}}\\
{{{{\bf \bar \Phi }}_u}}
\end{array}} \right] = I
\end{array} }
\end{align}
Moreover, if there exists a solution to this problem, then ${\bf{K}}={\bf \bar \Phi}_{u} {\bf \bar \Phi}_{x}^{-1}$ is mean-square stabilizing and achieves the desired closed-loop responses.  \vspace{3pt}

\noindent \textbf{Proof.} 
Let ${\bf{x}}={\bf{\bar x}}+{\bf{\tilde x}}$, where ${\bf{\bar x}}$ is the state solution of the nominal system \eqref{nom}, and ${\bf{\tilde x}}$ is the difference between the solution of the nominal system. Similarly, define ${\bf{u}}={\bf{\bar u}}+{\bf{\tilde u}}$. Using \eqref{ineqres}, one has $\|{\bf{\tilde x}}\|_{V_x}=\|{\bf{x}}-{\bf{\bar x}}\|_{V_x} \le 1$ and $\|{\bf{\tilde u}}\|_{V_u}=\|{\bf{u}}-{\bf{\bar u}}\|_{V_u} \le 1$. Using $\mathbf{\bar x}={\bf \bar \Phi _{{x}}} \mathbf{w}$ and $\mathbf{\bar u}={\bf \bar \Phi _{{u}}} \mathbf{w}$, and thus $\mathbf{x}={\bf \bar \Phi _{{x}}} \mathbf{w}+\mathbf{ \tilde x}$ and $\mathbf{u}={\bf \bar \Phi _{{u}}} \mathbf{w}+\mathbf{ \tilde u}$, the performance function in Problem P2 is transformed into the performance function in Problem P2. On the other hand, since the minimization decision variables are the closed-loop responses of the nominal system, the equality constraint imposes the response of the nominal system. 
On the other hand, if there exists a solution to this problem, then, based on \eqref{phix1} and \eqref{phiu1}, ${\bf{K}}={\bf \bar \Phi}_{u} {\bf \bar \Phi}_{x}^{-1}$. Besides, when the solution exists, the performance is bounded, which occurs if and only if $\text{lim}_{t \rightarrow \infty} \mathbb{E} \{x_t \, x_t^{\top} \}=0$. That is, the system is mean-square stable under the control gain and achieves the desired closed-loop responses, which completes the proof. \end{thm} \hfill $\square$ 

\section{system level synthesis for systems with multiplicative noise using scenario optimization}

In Problem P2, since the system response must satisfy the inequality constraint for all realizations of the uncertain system, it results in conservative and possible infeasible solutions.  To sidestep this difficulty, this section leverages sample-based approaches to provide solutions that are more tractable and less conservative by guaranteeing that the solution set will satisfy the inequality constraint for all but a small fraction of the systems in the family \eqref{mAB} and \eqref{mABB}. The key idea is that instead of maximizing over the worst-case realization of the set-valued system responses, several i.i.d samples of the system realizations will be made and an inequality will be imposed for each sample to assure that the closed-loop system responses found by the formulated optimization satisfy a set-valued map with a small radius for each realization.  Using scenario-based approach a bound on the number of samples will be provided to assure that their response will cover the entire ellipsoidal set described by the inequality in Problem P2, expect a small fraction of it, with high probability. To this end, we  present a probabilistic counterpart of Problem P2 and then leverage the scenario-based approach to efficiently solve it. 
 \vspace{6pt}

 Routine calculations show that
\begin{align}
\Scale[1]{{\bf{x}} = {({I_{nT}} - {{\cal Z}}({{\cal A}} + {{\cal B}}{{\bf{K}}}))^{ - 1}}{\bf{w}}},
\label{eq:15}
\end{align}
where 
\begin{align}
\Scale[1]{{{\cal A}} = \left[ {\begin{array}{*{20}{c}}
{{I_{T - 1}}}&{}\\
{}&0
\end{array}} \right] \otimes {A_0} + {\left[ {\begin{array}{*{20}{c}}
{\mathord{\buildrel{\lower3pt\hbox{$\scriptscriptstyle\frown$}} 
\over \Delta } }&{}\\
{}&0
\end{array}} \right]}{({I_T} \otimes {\bf A})} },
\label{eq:16}
\end{align}
\begin{align}
\Scale[1]{{{\cal B}} = \left[ {\begin{array}{*{20}{c}}
{{I_{T - 1}}}&{}\\
{}&0
\end{array}} \right] \otimes {B_0} + {\left[ {\begin{array}{*{20}{c}}
{\mathord{\buildrel{\lower3pt\hbox{$\scriptscriptstyle\frown$}} 
\over \Delta } }&{}\\
{}&0
\end{array}} \right]}{({I_T} \otimes {\bf B})} },
\label{eq:17}
\end{align}
\begin{align}
\Scale[1]{\mathord{\buildrel{\lower3pt\hbox{$\scriptscriptstyle\frown$}} 
\over \Delta }  = diag({\delta _0},{\delta _1},...,{\delta _{T - 1}}) \in {\mathbb{R} ^{(T-1)n \times (T-1)n{n_\delta }}} },
\label{eq:18}
\end{align}
and
\begin{align}
\Scale[1]{ {\delta _t}: = {\left[ {\begin{array}{*{20}{c}}
{\delta _t^{(1)}{I_{ n}}}& \cdots &{\delta _t^{({n_\delta })}{I_{ n}}}
\end{array}} \right]} \in {\mathbb{R} ^{ n{n_\delta }}}}.
\label{eq:33}
\end{align}

It follows from \eqref{eq:15}-\eqref{eq:18} that $\mathbf{x}={\bf \Phi _{{x}}} \mathbf{w}$ and $\mathbf{u}={\bf \Phi _{{u}}} \mathbf{w}$ with
\begin{align} 
&{{\bf{\Phi }}_{\bf{x}}} = {\left( {{I_{nT}} - {\cal Z}\left( {{A_0} + {B_0}{\bf{K}}} \right)} \right)^{ - 1}}{\left( {I + \bar \Delta } \right)^{ - 1}},\label{phix} \\
&{{\bf{\Phi }}_{\bf{u}}} = {\bf{K}}{\left( {{I_{nT}} - {\cal Z}\left( {{A_0} + {B_0}{\bf{K}}} \right)} \right)^{ - 1}}{\left( {I + \bar \Delta } \right)^{ - 1}}, \label{phiu}
\end{align} 
where 
\begin{align} 
\begin{array}{l}
\bar \Delta  =  - {\cal Z}\left[ {\begin{array}{*{20}{c}}
{\mathord{\buildrel{\lower3pt\hbox{$\scriptscriptstyle\frown$}} 
\over \Delta } }&{}\\
{}&0
\end{array}} \right]\left( {({I_T} \otimes {\bf{A}}) + ({I_T} \otimes {\bf{B}}){\bf{K}}} \right) \times \\
\,\,\,\,\,\,\,\,\,{({I_{nT}} - {\cal Z}\left( {{{\cal A}_0} + {{\cal B}_0}{\bf{K}}} \right))^{ - 1}},
\end{array} \label{eq:25}
\end{align}
with
${\bf{A}}: = {\left[ {\begin{array}{*{20}{l}}
{A_1^ \top }& \ldots &{A_{{n_\delta }}^ \top }
\end{array}} \right]^ \top } \in {\mathbb{R} ^{n{n_\delta } \times n}}$, and $\bf{B}: = {\left[ {\begin{array}{*{20}{l}}
{B_1^ \top }& \ldots &{B_{{n_\delta }}^ \top }
\end{array}} \right]^ \top } \in {\mathbb{R} ^{n{n_\delta } \times m}}$.




\begin{thm}\label{theorem:1}
Under the controller $u_t=K_t x_t$, the closed-loop system with multiplicative noise achieves the following set-valued system responses
\begin{align}
\Scale[1]{ \left[ {\begin{array}{*{20}{l}}
{\bf{x}}\\
{\bf{u}}
\end{array}} \right] = \left[\begin{array}{c}   {\bf \bar \Phi}_{x} \\  {\bf \bar \Phi}_{u}\end{array}\right]{\left( {{I_{nT}} + \bar \Delta } \right)^{ - 1}}{\bf{w}}},
\label{eq:101m}
\end{align}
Moreover, $\left\{{\bf \bar \Phi_{x}},{\bf \bar \Phi_{u}}\right\}$ satisfies
\begin{align}
\Scale[1]{\left[ {\begin{array}{*{20}{l}}
{I - {\cal Z}{{\cal A}_0}}&{ -  {\cal Z}{{\cal B}_0}}
\end{array}} \right]\left[\begin{array}{c}   {\bf \bar \Phi}_{x} \\ {\bf \bar \Phi}_{u}\end{array}\right] = {I_{nT}} + \bar \Delta  },
\label{eq:102m}
\end{align}
\end{thm} \vspace{5pt}
where $\bar \Delta$ is defined in \eqref{eq:25}. 

\noindent \textbf{Proof.} 
One has
\begin{align}
\Scale[1] {\begin{array}{l}
{\bf{K}} = {{{\bf{\bar \Phi }}}_{\bf{u}}}{{{\bf{\bar \Phi }}}_{\bf{x}}}^{ - 1} = {{{\bf{\bar \Phi }}}_{\bf{u}}}{({I_{nT}} + \bar \Delta )^{ - 1}}{\left( {{{{\bf{\bar \Phi }}}_{\bf{x}}}{{({I_{nT}} + \bar \Delta )}^{ - 1}}} \right)^{ - 1}}\\
\,\,\,\,\,\,\, = {{\bf{\Phi }}_{\bf{u}}}{{\bf{\Phi }}_{\bf{x}}}^{ - 1},
\end{array}}
\label{eq:103m}
\end{align}
where the last equality is obtained based on \eqref{phix} and \eqref{phiu} and definition of $\bar \Delta$. 
Now, using Proposition \ref{prop:1}, one has
\begin{align}
\Scale[1]{ \begin{array}{l}
\left[ {\begin{array}{*{20}{l}}
{I -  {\cal Z}{{\cal A}_0}}&{ -  {\cal Z}{{\cal B}_0}}
\end{array}} \right]\left[ {\begin{array}{*{20}{l}}
{\bf \bar \Phi_x}\\
{\bf \bar \Phi_u}
\end{array}} \right] {({I_{nT}} + {{\bar \Delta }})^{ - 1}} = {I_{nT}}
\end{array},}
\label{eq:105m}
\end{align}
since $\left(I+\Delta^{i, 0}\right)$ exists for $i=1, \ldots, T$, \eqref{eq:105m} is equivalent to \eqref{eq:102m}.
This completes the proof.\hfill $\square$

Now, recalling that $\mathord{\buildrel{\lower3pt\hbox{$\scriptscriptstyle\frown$}} \over \Delta }  = diag({\delta _0},{\delta _1},...,{\delta _{T - 1}})$
where ${\delta _t} = {\left[ {\begin{array}{*{20}{c}}
{\delta _t^{(1)}{I_{ n}}}& \cdots &{\delta _t^{({n_\delta })}{I_{ n}}}
\end{array}} \right]}$,
one can rewrite $\bar \Delta$ as
\begin{align}
\Scale[1]{
\bar \Delta  =  {{\mathscr{ R}}{\Theta }},
}
\label{eq:35}
\end{align}
where ${\Theta } = {\Theta _1}{\Theta _2} \in {\mathbb{R} ^{n{n_\delta }T \times nT}}$, 
\begin{align}
&{\Theta _1}: = \left( { - {{({I_{T}} \otimes {\bf{A}})}} - {{({I_{T}} \otimes \bf{B})}}{{\bf{K}}}} \right) \in {\mathbb{R} ^{n{n_\delta }T \times nT}}, 
\label{eq:36}
\\
&{\Theta _2}: = {\left( {{I_{nT}} - {{\cal Z}}\left( {{{\cal A}_0} + {{\cal B}_0}{{\bf{K}}}} \right)} \right)^{ - 1}} \in {\mathbb{R} ^{nT \times nT}}, 
\label{eq:37}
\end{align}
and 
\begin{align}
\Scale[0.9]{{\mathscr{ R}} := {{\cal Z}}{\left[ {\begin{array}{*{20}{c}}
{\mathord{\buildrel{\lower3pt\hbox{$\scriptscriptstyle\frown$}} 
\over \Delta } }&{}\\
{}&0
\end{array}} \right]} = {\left[ {\begin{array}{*{20}{c}}
0&{}&{}&{}\\
{{\delta _0}}&0&{}&{}\\
{}& \ddots &{}&{}\\
{}&{}&{{\delta _{T - 1}}}&0
\end{array}} \right].} }
\label{eq:34}
\end{align}
Note that $\bar \Delta$ is a random matrix since
${\mathscr{ R}}$ is a random matrix.

\begin{rem}
In Problem P2, the maximization over the closed-loop system error responses $\{{\bf{\tilde x}},{\bf{\tilde u}}\}$ assures that the performance is minimized over the worst-case realizable closed-loop responses. This problem can also be reformulated directly in terms of the error dynamics $\bar \Delta$ defined in \eqref{eq:25} as follows
\begin{align} 
\Scale[1]{\begin{array}{*{20}{l}}
{\mathop {{\rm{min}}}\limits_{{{{\bf{\bar \Phi }}}_x},{{{\bf{\bar \Phi }}}_u}} \mathop {\max }\limits_{\bar \Delta  \in {{\cal D}_{\bar \Delta }}} \left\| {\left[ {\begin{array}{*{20}{l}}
{{{\cal Q}^{1/2}}}&{}\\
{}&{{{\cal R}^{1/2}}}
\end{array}} \right]\left[ {\begin{array}{*{20}{l}}
{{{{\bf{\bar \Phi }}}_x}}\\
{{{{\bf{\bar \Phi }}}_u}}
\end{array}} \right]{{\left( {{I_{nT}} + \bar \Delta } \right)}^{ - 1}}{\bf{w}}} \right\|_F^2}\\
{{\mkern 1mu} s.t.{\mkern 1mu} {\mkern 1mu} {\mkern 1mu} \left[ {\begin{array}{*{20}{c}}
{\left( {I - {\cal Z}{{\cal A}_0}} \right)}&{ - {\cal Z}{{\cal B}_0}}
\end{array}{\mkern 1mu} {\mkern 1mu} {\mkern 1mu} {\mkern 1mu} } \right]\left[ {\begin{array}{*{20}{l}}
{{{{\bf{\bar \Phi }}}_x}}\\
{{{{\bf{\bar \Phi }}}_u}}
\end{array}} \right] = I}
\end{array} }
\end{align}
where ${{\cal D}_{\bar \Delta }}$ is the high-confidence bound of $\bar \Delta$, i.e., $\bar \Delta \in {{\cal D}_{\bar \Delta }}$ with probability at least $1-\epsilon$.

In both of these formulations, the optimization problems are bilevel optimizations. The next section shows how to turn the problem into a convex optimization problem by moving the uncertainties from the performance into the constraints. 
\end{rem}

The next theorem provides a probabilistic formulation of the SLS-based solution for the LQR control design of systems with multiplicative noise. The following proposition is required in the proof of the next theorem. \vspace{3pt}

\begin{prop}\label{Prop:3} \cite{Chen2007ANG}
For a random vector $X \in {\mathbb{R}}^{n}$ with cumulative distribution $\mathscr{F}(.)$,
\begin{align}
\Scale[1]{\mathbb{P}\{\| X-{\mathbb E}[X]|| \geq \varepsilon\} \leq \frac{\operatorname{{\mathbb V}ar}(X)}{\varepsilon^{2}}, \quad \forall \varepsilon>0, }
\label{eq:39}
\end{align}
where
\begin{align}
\Scale[0.9]{ {\mathbb V}ar(X): = {\rm{ }}\int_{\rho  \in {{\mathbb{R}}^n}} {{{\left\| {\rho  - {\mathbb E}[X]} \right\|}^2}} d{\mathscr{F}}(\rho ). }
\label{eq:44P2}
\end{align}
\end{prop} \vspace{6pt}

\begin{thm}\label{theorem:2}
Under the controller $u_t=K_t x_t$, the set-valued closed-loop responses of the system with multiplicative noise satisfy the chance constraint
\begin{align}
\Scale[0.85]{\begin{array}{l}
{\mathbb P} \left\{ {\left\| {\text{vec}\left\{ {\left[ {\begin{array}{*{20}{l}}
{I - {\cal Z}{{\cal A}_0}}&{ - {\cal Z}{{\cal B}_0}}
\end{array}} \right]\left[\begin{array}{c}   {\bf \bar \Phi}_{x} \\ {\bf \bar \Phi}_{u}\end{array}\right] - {I}} \right\}  } \right\| \ge \varepsilon } \right\} \le 
{{\lambda \mathord{\left/
 {\vphantom {\lambda {{\varepsilon ^2}}}} \right.
 \kern-\nulldelimiterspace} {{\varepsilon ^2}}}}  ,
\end{array} }
\label{eq:41}
\end{align}
$\forall \varepsilon  > 0$, where 
\begin{align}
\Scale[0.9]{\lambda := {\mathbb V}ar\left\{ {\left( {{{ { {\Theta }} }^ \top } \otimes {I_{Tn }}} \right){{\mathscr R}_{\text{vec}}}} \right\}\mathop. }
\label{eq:44}
\end{align}
with ${{\mathscr R}_{\text{vec}}}:=\text{vec}({\mathscr R})$.
\end{thm}
\textbf{Proof.} 
Note that ${\mathop{\rm vec}\nolimits} (ABC) = \left( {{C^ \top } \otimes A} \right){\mathop{\rm vec}\nolimits} (B)$ and ${\mathbb E}[\left( {{{ { {\Theta }} }^ \top } \otimes {I}} \right) {{\mathscr R}_{\text{vec}}}]=0$. Therefore,
\begin{align}
\Scale[0.9]{{\mathop{\rm vec}\left\{ {\left[ {\begin{array}{*{20}{l}}
{I - {\cal Z}{{\cal A}_0}}&{ - {\cal Z}{{\cal B}_0}}
\end{array}} \right]\left[\begin{array}{c}   {\bf \bar \Phi}_{x} \\  {\bf \bar \Phi}_{u}\end{array}\right] - {I}} \right\} = \left( {{{ { {\Theta }} }^ \top } \otimes {I}} \right) {{\mathscr R}_{\text{vec}}} }. }
\label{eq:42}
\end{align}
Invoking Proposition \ref{Prop:3}, one has
\begin{align}
\Scale[1]{{\mathbb {P}}\left\{ {\left\| {\left( {{{{ {\Theta }} }^ \top } \otimes {I}} \right){{\mathscr R}_{\text{vec}}} } \right\|| \ge \varepsilon } \right\} \le \frac{{{\mathbb V}ar\left\{ {\left( {{{ { {\Theta }} }^ \top } \otimes {I}} \right){{\mathscr R}_{vec}}} \right\}}}{{{\varepsilon ^2}}}, }
\label{eq:45}
\end{align}
$\forall \varepsilon  > 0$, which implies \eqref{eq:41}.
This completes the proof.   \hfill $\square$ \vspace{6pt}

The following formulation leverages the results of Theorem 4 to reformulate Problem P2 as a chance-constraint optimization problem. \vspace{6pt}

For a given risk level $\varepsilon$, the chance-constrained formulation of Problem P1 becomes
\begin{align}
\Scale[0.85]{ \text{P} \, 3: \begin{array}{*{20}{l}}
{\mathop {{\rm{min}}}\limits_{\bar \Phi_x, \bar \Phi_u} {{\left\| {\left[ {\begin{array}{*{20}{c}}
{{{\cal Q}^{1/2}}}&{}\\
{}&{{{\cal R}^{1/2}}}
\end{array}} \right]\left[ {\begin{array}{*{20}{l}}
{\bf \bar \Phi}_{x} \\  {\bf \bar \Phi}_{u}
\end{array}} \right]{{\left( {I_{nT } + \bar \Delta (\delta )} \right)}^{ - 1}}{\bf w}} \right\|}_F^2}}\\
{{\rm{s}}{\rm{.t}}{\rm{. }}\, \mathbb{P} \left\{ {\left\| {\mathop{\rm vec}\left\{ {\left[ {\begin{array}{*{20}{l}}
{I - {\cal Z}{{\cal A}_0}}&{ - {\cal Z}{{\cal B}_0}}
\end{array}} \right]\left[ {\begin{array}{*{20}{l}}
{\bf \bar \Phi}_{x} \\  {\bf \bar \Phi}_{u}
\end{array}} \right] - {I}} \right\}} \right\| \ge \varepsilon } \right\} \le  {\lambda  \mathord{\left/
 {\vphantom {\lambda  {{\varepsilon ^2}}}} \right.
 \kern-\nulldelimiterspace} {{\varepsilon ^2}}} ,}
\end{array}}         \label{eq:15*}
\end{align}
or equivalently,
\begin{align}
\Scale[0.82]{  \begin{array}{*{20}{l}}
{\mathop {{\rm{min}}}\limits_{{\bf \Phi _x},{\bf \Phi _u}} {{\left\| {\left[ {\begin{array}{*{20}{c}}
{{{\cal Q}^{1/2}}}&{}\\
{}&{{{\cal R}^{1/2}}}
\end{array}} \right]\left[ {\begin{array}{*{20}{l}}
{{\bf \Phi _x}}\\
{{\bf \Phi _u}}
\end{array}} \right]{\bf w}} \right\|}_F^2}}\\
{\mathbb{P} {\mkern 1mu} \left\{ {\left\| {\mathop{\rm vec}\{ {\left[ {\begin{array}{*{20}{l}}
{I - {\cal Z}{{\cal A}_0}}&{ - {\cal Z}{{\cal B}_0}}
\end{array}} \right]\left[ {\begin{array}{*{20}{l}}
{{\bf \Phi _x}}\\
{{\bf \Phi _u}}
\end{array}} \right]\left( {{I} + \bar \Delta } \right) - {I}} \}} \right\| \le \varepsilon } \right\} \ge  1- {\lambda  \mathord{\left/
 {\vphantom {\lambda  {{\varepsilon ^2}}}} \right.
 \kern-\nulldelimiterspace} {{\varepsilon ^2}}} }
\end{array}}
\label{eq:48**}
\end{align}
Note that
\begin{align}
\Scale[1]{ \left[ {\begin{array}{*{20}{l}}
{{\bf \bar \Phi _x}}\\
{{\bf \bar \Phi _u}}
\end{array}} \right] = \left[ {\begin{array}{*{20}{l}}
{{\bf \Phi _x}}\\
{{\bf \Phi _u}}
\end{array}} \right]\left( {{I_{nT}} + \bar \Delta } \right),}
\label{eq:48**}
\end{align}
which makes \eqref{eq:15*} and \eqref{eq:48**} the same. \vspace{6pt}

Note that in contrast to the robust formulation in Problem P2, the constraint is over the entire high-probability system responses and not just the nominal systems, as the maximization over the model uncertainty is removed. This formulation paves the way to provide tractable solutions of the optimal control of systems with multiplicative noise. 


\begin{cor} Problem P3 can be reformulated as
\begin{align}
\Scale[1]{  \text{P} \, 4: \begin{array}{*{20}{l}}
{\mathop {{\rm{min}}}\limits_{\bf{{ {\Phi }}_x},{{\bf {\Phi }}_u}} {{\left\| {\left[ {\begin{array}{*{20}{c}}
{{{\cal Q}^{1/2}}}&{}\\
{}&{{{\cal R}^{1/2}}}
\end{array}} \right]\left[ {\begin{array}{*{20}{l}}
{{\bf{{\Phi }}_x}}\\
{{\bf{{\Phi }}_u}}
\end{array}} \right]{\bf w}} \right\|}_F^2}}\\
{s.t.\,{\mkern 1mu} \mathbb{P}\left\{ {\Psi  \le \frac{{2\varepsilon }}{{\sqrt {{n^2}{T^2}} }}{\bf{\mathord{\buildrel{\lower3pt\hbox{$\scriptscriptstyle\leftrightarrow$}} 
\over 1} }}_{nT }^{L - tri}} \right\} \ge 1 -  {\lambda  \mathord{\left/
 {\vphantom {\lambda  {{\varepsilon ^2}}}} \right.
 \kern-\nulldelimiterspace} {{\varepsilon ^2}}} }
\end{array} }
\label{eq:Problem5}
\end{align}
where 
\begin{align} 
\Psi: ={\left[ {\begin{array}{*{20}{c}}
{\Phi _x^{0,0} - I}&{}&{}&{}\\
{{\Lambda _{1,2}}}&{\Phi _x^{1,0} - I}&{}&{}\\
 \vdots & \ddots & \ddots &{}\\
{{\Lambda _{1,T}}}& \cdots &{{\Lambda _{T - 1,T}}}&{\Phi _x^{T,0} - I}
\end{array}} \right], } \label{eq:Psi}
\end{align}
with 
\begin{align} \label{lambda} {\Lambda _{i,j}}: = \left[ {\begin{array}{*{20}{c}}
I&{\left[ {\Omega (\Delta )} \right]_i^c}
\end{array}} \right]\left[ {\begin{array}{*{20}{c}}
{\left[ {\Upsilon (\Phi )} \right]_{i,j}^{cr}}\\
{\left[ {\Upsilon (\Phi )} \right]_j^r}
\end{array}} \right],
\end{align}
\begin{align}
\Scale[0.75]{ \Omega (\Delta ) := \left[ {\begin{array}{*{20}{c}}
I&{}&{}&{}\\
{{\Delta ^{0,0}}}&I&{}&{}\\
 \vdots & \ddots & \ddots &{}\\
{{\Delta ^{T - 1,T - 1}}}& \cdots &{{\Delta ^{T - 1,1}}}&I
\end{array}} \right] = \left[ {\begin{array}{*{20}{c}}
0&{}&{}&{}\\
{{\delta _0}}&0&{}&{}\\
{}& \ddots &{}&{}\\
{}&{}&{{\delta _{T - 1}}}&0
\end{array}} \right]\Theta  + I,}
\end{align}
and
\begin{align}
\Scale[0.9]{\Upsilon (\Phi ) := \left[ {\begin{array}{*{20}{c}}
{\Phi _x^{0,0}}&{}&{}&{}\\
{{\Pi _{1,1}}}&{\Phi _x^{1,0}}&{}&{}\\
 \vdots & \ddots & \ddots &{}\\
{{\Pi _{T,T}}}& \cdots &{{\Pi _{T,1}}}&{\Phi _x^{T,0}}
\end{array}} \right],}
\end{align}
where ${\Pi _{k,m}} = \Phi _x^{k,m} - {A_0}\Phi _x^{k - 1,m - 1} - {B_0}\Phi _u^{k - 1,m - 1}$ and where $\mathord{\buildrel{\lower3pt\hbox{$\scriptscriptstyle\leftrightarrow$}} 
\over {\bf 1}} _{nT }^{L - tri}$  denotes the ${nT \times nT}$-dimensional low-triangular matrix whose components are all one.

\noindent \textbf{Proof.}
Noting
\begin{align}
\Scale[0.9]{{\left[ {\begin{array}{*{20}{c}}
0&{}&{}&{}\\
{{\delta _0}}&0&{}&{}\\
{}& \ddots &{}&{}\\
{}&{}&{{\delta _{T - 1}}}&0
\end{array}} \right]}{\Theta } = {\left[ {\begin{array}{*{20}{c}}
0&{}&{}&{}\\
{{\Delta ^{0,0}}}&0&{}&{}\\
 \vdots & \ddots & \ddots &{}\\
{{\Delta ^{T - 1,T - 1}}}& \cdots &{{\Delta ^{T - 1,1}}}&0
\end{array}} \right],} }
\end{align}
and after some manipulation, one has
\begin{align}
{\begin{array}{*{20}{l}}
{\left[ {\begin{array}{*{20}{l}}
{I - {\cal Z}{{\cal A}_0}}&{ - {\cal Z}{{\cal B}_0}}
\end{array}} \right]\left[ {\begin{array}{*{20}{l}}
{{\bf{{\Phi }}_x}}\\
{{\bf{{\Phi }}_u}}
\end{array}} \right]\left( {I + \bar \Delta {\mkern 1mu} {\mkern 1mu} } \right) - I = } \Psi,
\end{array}}
\end{align}
where $\Psi$ is given \eqref{eq:Psi}. 
Now, one can show that
\begin{align}
\Scale[0.9]{\left[ {\begin{array}{*{20}{c}}
{\Phi _x^{0,0} - I}&{}&{}&{}\\
{{\Lambda _{1,2}}}&{\Phi _x^{1,0} - I}&{}&{}\\
 \vdots & \ddots & \ddots &{}\\
{{\Lambda _{1,T}}}& \cdots &{{\Lambda _{T - 1,T}}}&{\Phi _x^{T,0} - I}
\end{array}} \right] \leq \frac{{2\varepsilon }}{{\sqrt {{n^2}{T^2}} }}{\bf{\mathord{\buildrel{\lower3pt\hbox{$\scriptscriptstyle\leftrightarrow$}} 
\over 1} }}_{nT \times nT}^{L - tri}},
\label{eq:P4toP5}
\end{align}
which  implies 
\begin{align}
\Scale[0.9]{ \left\| {\mathop{\rm vec}\left\{ {\left[ {\begin{array}{*{20}{l}}
{I - {\cal Z}{{\cal A}_0}}&{ - {\cal Z}{{\cal B}_0}}
\end{array}} \right]\left[ {\begin{array}{*{20}{l}}
{{\bf {{\Phi }}_x}}\\
{{\bf {{\Phi }}_u}}
\end{array}} \right]\left( {I + \bar \Delta {\mkern 1mu} {\mkern 1mu} } \right) - I} \right\}} \right\| \le \varepsilon }.
\label{eq:1qqq}
\end{align}
This completes the proof. \hfill $\square$
\end{cor}

A safe solution to this joint chance constraint problem can be found by solving
\begin{align}
\Scale[0.92]{\text{P} \, 5:   
\begin{array}{*{20}{l}}
{\mathop {{\rm{min}}}\limits_{{{{\Phi }}_x},{{{\Phi }}_u}} {{\left\| {\left[ {\begin{array}{*{20}{c}}
{{{\cal Q}^{1/2}}}&{}\\
{}&{{{\cal R}^{1/2}}}
\end{array}} \right]\left[ {\begin{array}{*{20}{l}}
{{\bf{{\Phi }}_x}}\\
{{\bf {{\Phi }}_u}}
\end{array}} \right]{\bf w}} \right\|}_F^2}}\\
{s.t.\,{\mkern 1mu} \mathbb{P}\left\{ {{\Lambda _{i,j}} \le \frac{{2\varepsilon }}{{\sqrt {{n^2}{T^2}} }}{\bf{\mathord{\buildrel{\lower3pt\hbox{$\scriptscriptstyle\leftrightarrow$}} 
\over 1} }},{\mkern 1mu} {\mkern 1mu} i \ge j,\,{\mkern 1mu} {\mkern 1mu} i,{\mkern 1mu} j = 1,...,T} \right\} \ge 1 -  {\lambda  \mathord{\left/
 {\vphantom {\lambda  {{\varepsilon ^2}}}} \right.
 \kern-\nulldelimiterspace} {{\varepsilon ^2}}}, }
\end{array}}
\label{eq:Problem6}
\end{align}
where $\Lambda _{i,j}$ is defined in \eqref{lambda}.
We assume, for simplicity, that $\Delta ^{i,j} \in {\Delta}_s, \forall i,j= 0,...,T$. 

Note that the uncertain constraints in Problem $\text{P} \, 7$ are linear inequalities, but involves an infinite number of constraints, since $\Delta$ is uncountable. 
In the so-called scenario approach, which is a data-driven relaxation of Problem $\text{P} \, 7$,
 a finite (say, $N$-dimensional) set of system realizations is sampled. 
Note that this approach results in a less conservative solution compared to robust approach. To this end, $\text{P} \, 7$ should be first rephrased in epigraph form \cite{calafiore2014optimization} as
\begin{align}
\Scale[0.95]{\text{P} \, 6: \begin{array}{*{20}{l}}
{\mathop {{\rm{min}}}\limits_{{{{\Phi }}_x},{{{\Phi }}_u},{\kern 1pt} \alpha } {\mkern 1mu} {\mkern 1mu} \alpha {\mkern 1mu} }\\
{\begin{array}{*{20}{l}}
{s.t.{\mkern 1mu} \,{\mkern 1mu} {{\left\| {\left[ {\begin{array}{*{20}{c}}
{{{\cal Q}^{1/2}}}&{}\\
{}&{{{\cal R}^{1/2}}}
\end{array}} \right]\left[ {\begin{array}{*{20}{l}}
{{\bf {{\Phi }}_x}}\\
{{\bf {{\Phi }}_u}}
\end{array}} \right]{\bf w}} \right\|}_F^2} \le \alpha {\mkern 1mu} {\mkern 1mu} }\\
\quad  \mathbb{P} \, {\left\{ {{\Lambda _{i,j}} \le \frac{{2\varepsilon }}{{\sqrt {{n^2}{T^2}} }}{\bf{\mathord{\buildrel{\lower3pt\hbox{$\scriptscriptstyle\leftrightarrow$}} 
\over 1} }},{\mkern 1mu} {\mkern 1mu} {\mkern 1mu} i \ge j,\,{\mkern 1mu} {\mkern 1mu} i,{\mkern 1mu} j = 1,...,T} \right\} \ge 1 -  {\lambda  \mathord{\left/
 {\vphantom {\lambda  {{\varepsilon ^2}}}} \right.
 \kern-\nulldelimiterspace} {{\varepsilon ^2}}} }\\
\quad {\forall {\Delta ^{i,j}} \in {\Delta _s},{\mkern 1mu} {\rm{i}},{\rm{j = 0}},...,{\rm{T - 1}}}
\end{array}}
\end{array}}
\label{eq:Problem8}
\end{align}
Now, the main idea is to replace the optimization Problem $\text{P} \, 8$ by its following sampled-based counterpart (the scenario-based problem)
\begin{align}
\Scale[1]{\text{P} \, 7: \begin{array}{*{20}{l}}
{{\bf {{\Phi }}_x}^{SC},{\bf {{\Phi }}_u}^{SC} = \mathop {{\rm{arg}}{\mkern 1mu} {\rm{min}}}\limits_{{{{\Phi }}_x},{{{\Phi }}_u},{\kern 1pt} \alpha } {\mkern 1mu} {\mkern 1mu} \alpha {\mkern 1mu} }\\
{\begin{array}{*{20}{l}}
{s.t.{\mkern 1mu} {\mkern 1mu} {{\left\| {\left[ {\begin{array}{*{20}{c}}
{{{\cal Q}^{1/2}}}&{}\\
{}&{{{\cal R}^{1/2}}}
\end{array}} \right]\left[ {\begin{array}{*{20}{l}}
{{\bf{{\Phi }}_x}}\\
{{\bf {{\Phi }}_u}}
\end{array}} \right]{\bf w}} \right\|}_F^2} \le \alpha {\mkern 1mu} {\mkern 1mu} }\\
\quad {{\Lambda _{i,j}} \le \frac{{2\varepsilon }}{{\sqrt {{n^2}{T^2}} }}{\bf{\mathord{\buildrel{\lower3pt\hbox{$\scriptscriptstyle\leftrightarrow$}} 
\over 1} }},{\mkern 1mu} {\mkern 1mu} {\mkern 1mu} i \ge j,\,{\mkern 1mu} {\mkern 1mu} i,{\mkern 1mu} j = 1,...,T}\\
\quad
{\forall k = 1,...,N}
\end{array}}
\end{array}}
\label{eq:Problem9}
\end{align}
where ${\Delta ^k}$, $k=1,...,N$  are i.i.d. samples extracted. Moreover, the near-optimal solutions ${{{{\bf \Phi }}_x}}^{\star SC}$ and ${{{{\bf \Phi }}_u}}^{\star SC}$ of this optimization problem
are random variables that depend on the random extractions of the system's uncertainties ${\Delta ^1}, . . . , {\Delta ^N}$. 

\begin{rem}\label{Remark:2}
It is worth noting that $\text{P} \, 9$ is now a  convex optimization problem with a finite number of constraints, and consequently, it is efficiently solvable.
\end{rem}

The following standard assumption is routinely made in the literature on the scenario approach \cite{Campi_Garatti_2018,Calafiore2011ResearchOP,Calafiore2017RepetitiveSD}. \vspace{3pt}
\begin{assm}\label{Assum:1}
$\forall N \geq N_{SC}^{\star}$, $\text{P} \, 9$ is feasible and attains a unique optimal solution ${\bf \Phi}_{SC}^\star$.
\end{assm} \vspace{3pt}

\begin{thm}\label{theorem:4}
Under Assumption 1, $\text{P} \, 9$  is feasible and attains a unique optimal solution ${\bf \Phi}_{SC}^\star := \{{{{{\bf \Phi }}_x}}^{\star SC}, {{{{\bf \Phi }}_u}}^{\star SC}\}$. Given $\beta \in (0,1)$, if the number of scenarios $N$ satisfies the relation
\begin{align}
\Scale[1]{N \geq N_{SC}^{\star}:=\left[\frac{2}{1-\epsilon}\left(\log \frac{1}{\beta}+n(m+n)T^2\right)\right] ,}
\label{eq:theorem:4}
\end{align}
then ${\bf \Phi}_{SC}^\star$ satisfies the chance-constrained program \eqref{eq:Problem9} with confidence $(1-\beta)$.
\end{thm} \vspace{3pt}
 \textbf{Proof}. 
The proof follows from the key results in \cite{Calafiore2006TheSA} (i.e., Theorem 1 and Corollary 1) and \cite{Calafiore2009OnTE} (i.e., Proposition 2.1).   \hfill $\square$ \vspace{3pt}

\begin{rem}\label{Remark:3}
Note that Theorem \ref{theorem:4} states that the solution ${\bf \Phi}_{SC}^\star$ is feasible for all the constraints in P7 with high probability $(1-\beta)$, except possibly for those in a set having probability measure smaller than $\epsilon \in (0,1) $ \cite{Calafiore2010RandomCP}. 
Despite any probability distribution on the uncertainties, Theorem \ref{theorem:4} provides a promising tool  to compute a sufficient number of scenarios $ N_{SC}^{\star}$  a priori, before any constraint is extracted, that guarantees a certain level of robustness. 
In practice, the $\beta $ can be fixed to a very small value (say, ${10}^{-6}$), without increasing too much the required number of scenarios \cite{Campi_Garatti_2018}. Note that in contrast to some applications of scenario approach for which it is hard to collect i.i.d samples (example i.i.d samples of the system's states), here, there is no implication inn collecting i.i.d samples since the multiplicative dynamics are known. 
\end{rem}

\section{Simulation}

\begin{figure*}[pt]
     \centering
     \begin{subfigure}[b]{0.3\textwidth}
         \centering
         \includegraphics[width=\textwidth]{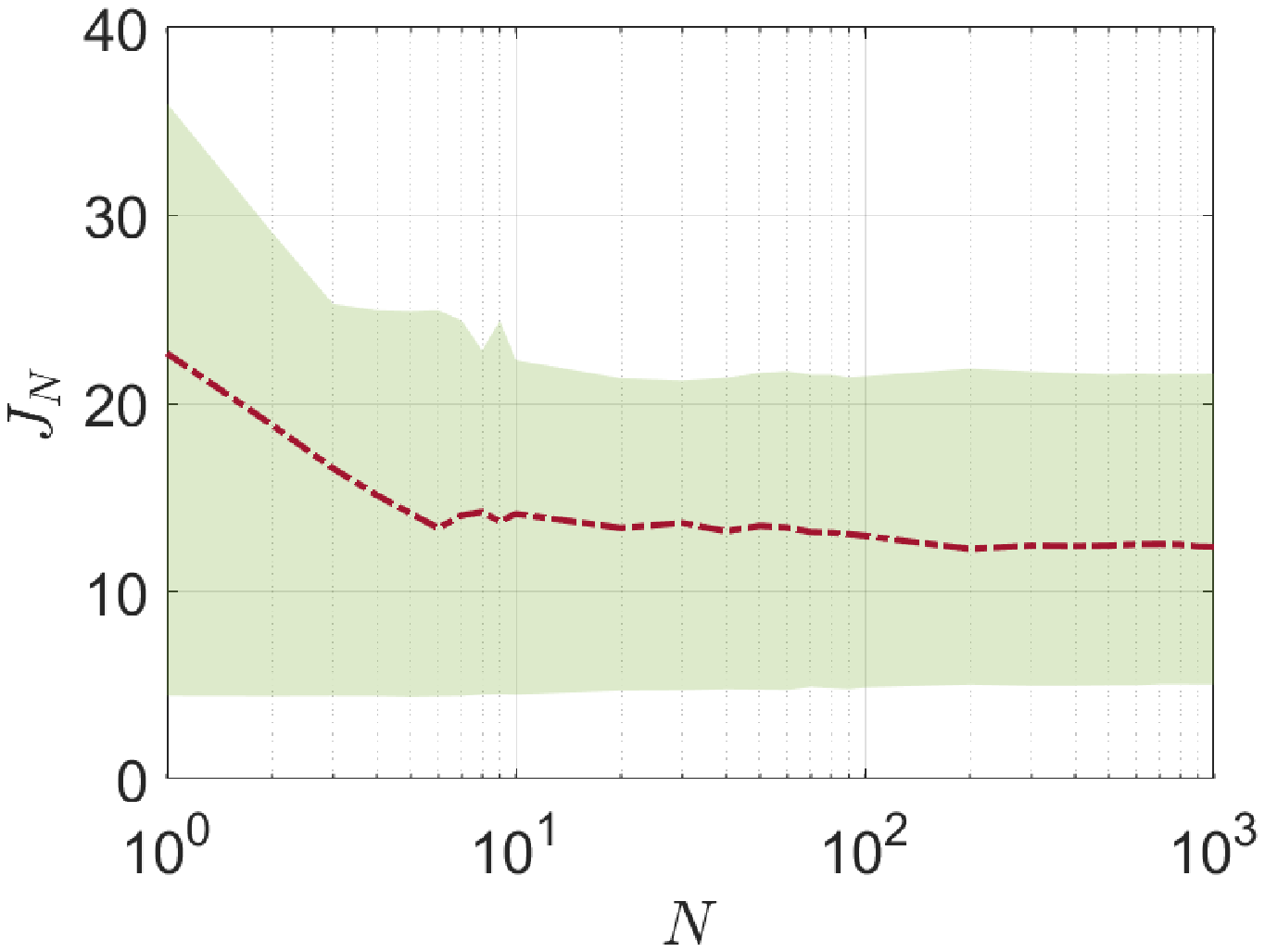}
         \caption{}
         \label{fig1}
     \end{subfigure}
     \hfill
     \begin{subfigure}[b]{0.3\textwidth}
         \centering
         \includegraphics[width=\textwidth]{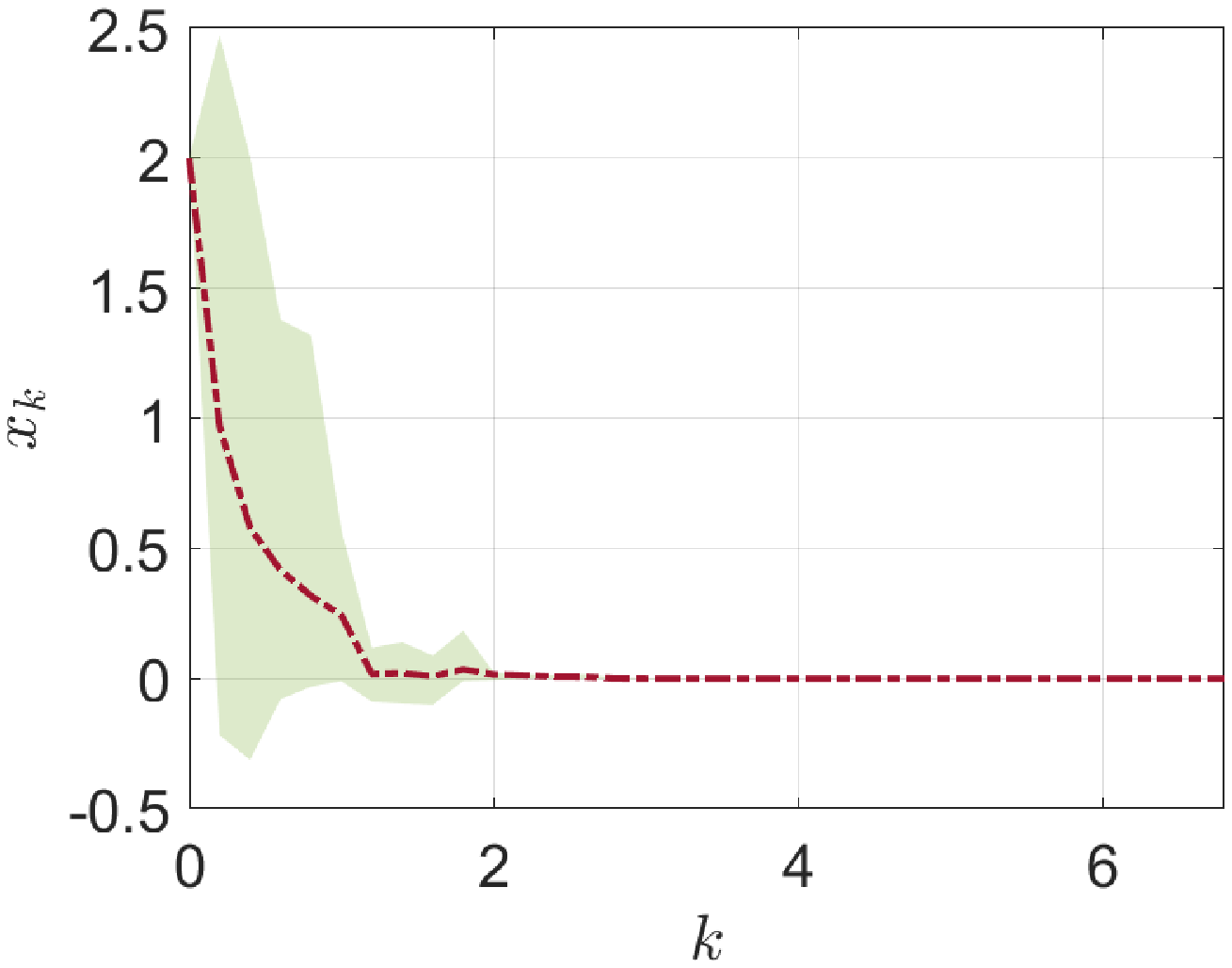}
         \caption{}
         \label{fig2}
     \end{subfigure}
     \hfill
     \begin{subfigure}[b]{0.3\textwidth}
         \centering
         \includegraphics[width=\textwidth]{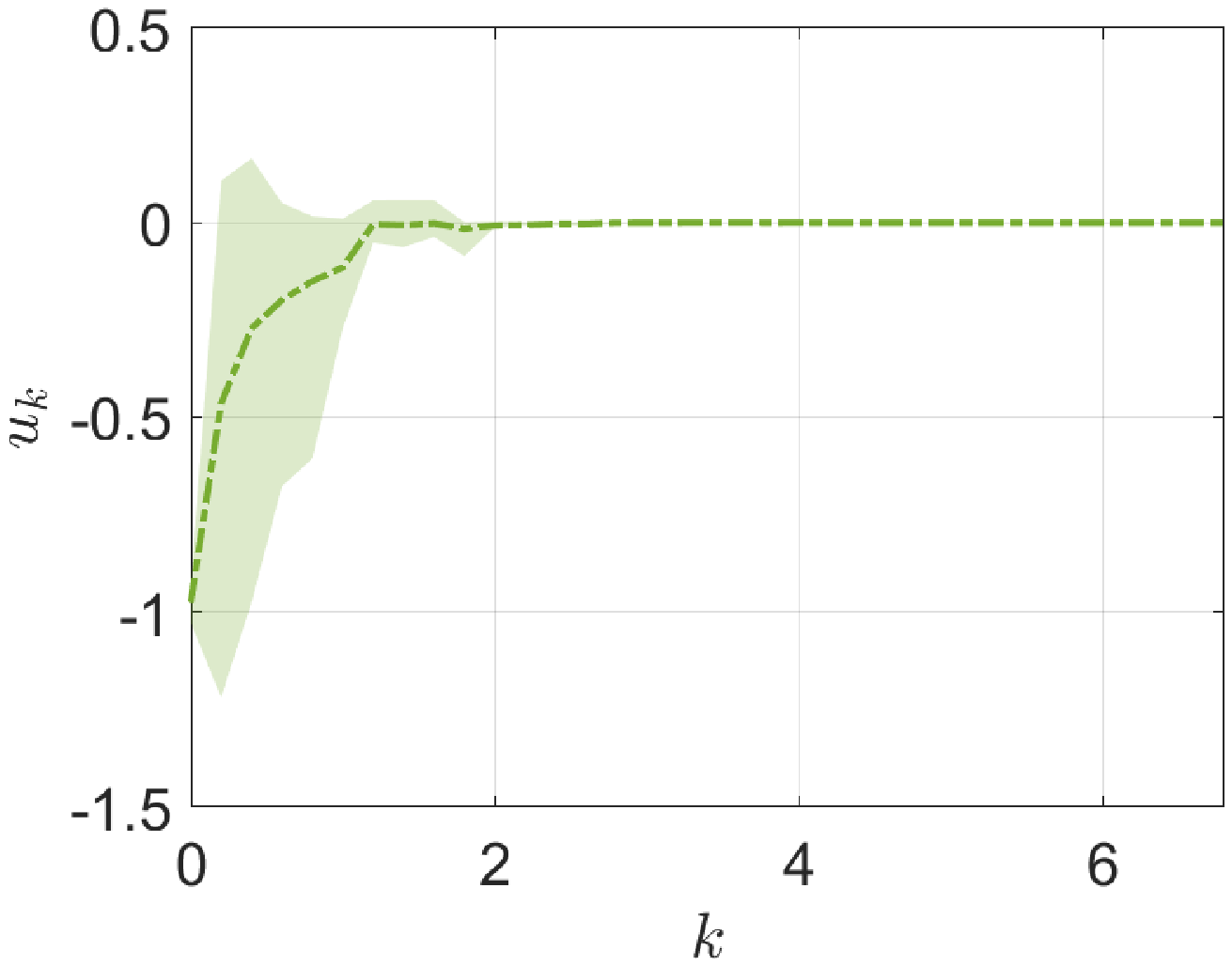}
         \caption{}
         \label{fig3}
     \end{subfigure}
        \caption{The performance, state, and control trajectories. (a) The performance function versus the number of scenarios $N$, (b)   system state trajectory, and (c) control input $u$.
        The shaded area and dashed line represent the results between $[10 \%, 90 \%]$ quantiles and the means across $25$ independent experiments.}
        \label{figK}
\end{figure*}

The efficiency of the proposed algorithm is verified using the following linear dynamical system given by
\begin{align}
{x_{t + 1}} = (0.8+ {\varepsilon _t}){x_t} + 0.5{u_t},
\label{ex1}
\end{align}
and the cost function is assumed to be  $J = \sum\limits_{t = 0}^{\top}  {({x_t^2} + {u_t^2})}$.
  The disturbances  ${\left\{ {{\varepsilon _t}} \right\}_{t \in {1,...,T}}}$  are i.i.d. random variables generated by a truncated normal distribution with known parameters $\mu  = 0$  and $\sigma  = 0.5$. 

 The trajectory of state and the designed control input for the case of $1000$ scenario samples are displayed in Figs.~\ref{fig2} and ~\ref{fig3}, respectively. Fig.~\ref{fig1} shows the performance as a function of the number of scenarios. One can see from Fig.~\ref{fig1} that by choosing a greater number for scenario samples, the mean and variance of the performance will be diminished significantly and, therefore,  reach a better performance.    




\section{Conclusion}
 In this paper, we developed a sampled-based convex optimization approach for solving the closed-loop state-feedback control of a discrete-time LQR problem for systems affected by multiplicative noise. To synthesize a tractable state-feedback policy, first, we leveraged the SLS framework to reformulate the problem as a min-max optimization over entire set-valued closed-loop system responses. Then, we reformulated this optimization problem as a chance-constrained program (CCP) in which stochastic guarantees are provided for all, but a small fraction of possible closed-loop system responses. To approximately solve the CCP without the requirement of knowing the probabilistic description of the uncertainty in the system matrices, we utilized the scenario approach, which provides probabilistic guarantees based on a finite number of system's realizations and results in a convex optimization program with moderate computational complexity. Finally, numerical simulations were presented to illustrate the theoretical findings.

\bibliographystyle{IEEEtran}
\bibliography{IEEEabrv,references.bib}

\begin{thebibliography}{10}
\providecommand{\url}[1]{#1}
\csname url@samestyle\endcsname
\providecommand{\newblock}{\relax}
\providecommand{\bibinfo}[2]{#2}
\providecommand{\BIBentrySTDinterwordspacing}{\spaceskip=0pt\relax}
\providecommand{\BIBentryALTinterwordstretchfactor}{4}
\providecommand{\BIBentryALTinterwordspacing}{\spaceskip=\fontdimen2\font plus
\BIBentryALTinterwordstretchfactor\fontdimen3\font minus
  \fontdimen4\font\relax}
\providecommand{\BIBforeignlanguage}[2]{{%
\expandafter\ifx\csname l@#1\endcsname\relax
\typeout{** WARNING: IEEEtran.bst: No hyphenation pattern has been}%
\typeout{** loaded for the language `#1'. Using the pattern for}%
\typeout{** the default language instead.}%
\else
\language=\csname l@#1\endcsname
\fi
#2}}
\providecommand{\BIBdecl}{\relax}
\BIBdecl

\bibitem{kalman1960contributions}
R.~E. Kalman \emph{et~al.}, ``Contributions to the theory of optimal control,''
  \emph{Bol. soc. mat. mexicana}, vol.~5, no.~2, pp. 102--119, 1960.

\bibitem{anderson1990optimal}
B.~D. Anderson and J.~B. Moore, ``Optimal control: linear quadratic methods,''
  1990.

\bibitem{doyle1978guaranteed}
J.~C. Doyle, ``Guaranteed margins for {LQG} regulators,'' \emph{{IEEE} Trans.
  Automat. Contr.}, vol.~23, no.~4, pp. 756--757, 1978.

\bibitem{scampicchio2021stable}
A.~Scampicchio, A.~Aravkin, and G.~Pillonetto, ``Stable and robust lqr design
  via scenario approach,'' \emph{Automatica}, vol. 129, p. 109571, 2021.

\bibitem{anderson2019system}
J.~Anderson, J.~C. Doyle, S.~H. Low, and N.~Matni, ``System level synthesis,''
  \emph{Annu. Rev. Control}, vol.~47, pp. 364--393, 2019.

\bibitem{xue2021data}
A.~Xue and N.~Matni, ``Data-driven system level synthesis,'' in \emph{Learning
  for Dynamics and Control}.\hskip 1em plus 0.5em minus 0.4em\relax PMLR, 2021,
  pp. 189--200.

\bibitem{bemporad1999robust}
A.~Bemporad and M.~Morari, ``Robust model predictive control: A survey,'' in
  \emph{Robustness in identification and control}.\hskip 1em plus 0.5em minus
  0.4em\relax Springer, 1999, pp. 207--226.

\bibitem{raimondo2009min}
D.~M. Raimondo, D.~Limon, M.~Lazar, L.~Magni, and E.~F. ndez Camacho, ``Min-max
  model predictive control of nonlinear systems: A unifying overview on
  stability,'' \emph{Eur. J. Control}, vol.~15, no.~1, pp. 5--21, 2009.

\bibitem{saltik2018outlook}
M.~B. Salt{\i}k, L.~{\"O}zkan, J.~H. Ludlage, S.~Weiland, and P.~M. Van~den
  Hof, ``An outlook on robust model predictive control algorithms: Reflections
  on performance and computational aspects,'' \emph{J. Process Control},
  vol.~61, pp. 77--102, 2018.

\bibitem{LQRdata1}
C.~{De Persis} and P.~Tesi, ``Low-complexity learning of linear quadratic
  regulators from noisy data,'' \emph{Automatica}, vol. 128, p. 109548, 2021.

\bibitem{LQRdata2}
S.~Dean, H.~Mania, N.~Matni, B.~Recht, and S.~Tu, ``Regret bounds for robust
  adaptive control of the linear quadratic regulator,''
  \emph{arXiv:1805.09388}, 2018.

\bibitem{LQRdata3}
S.~Tu, ``Sample complexity bounds for the linear quadratic regulator,'' Ph.D.
  dissertation, EECS Department, University of California, Berkeley, May 2019.

\bibitem{gravell2019learning}
B.~Gravell, P.~M. Esfahani, and T.~Summers, ``Learning robust control for lqr
  systems with multiplicative noise via policy gradient,'' \emph{arXiv,
  arXiv:1905.13547}, 2019.

\bibitem{coppens2020data}
P.~Coppens, M.~Schuurmans, and P.~Patrinos, ``Data-driven distributionally
  robust lqr with multiplicative noise,'' in \emph{Learning for Dynamics and
  Control}.\hskip 1em plus 0.5em minus 0.4em\relax PMLR, 2020, pp. 521--530.

\bibitem{gravell2020robust}
B.~J. Gravell, P.~M. Esfahani, and T.~H. Summers, ``Robust control design for
  linear systems via multiplicative noise,'' \emph{IFAC-PapersOnLine}, vol.~53,
  no.~2, pp. 7392--7399, 2020.

\bibitem{pang2022robust}
B.~Pang and Z.-P. Jiang, ``Robust reinforcement learning for stochastic linear
  quadratic control with multiplicative noise,'' \emph{Trends in Nonlinear and
  Adaptive Control}, pp. 249--277, 2022.

\bibitem{jongeneel2019robust}
W.~Jongeneel, T.~Summers, and P.~M. Esfahani, ``Robust linear quadratic
  regulator: Exact tractable reformulation,'' in \emph{Proc. IEEE Conf. Decis.
  Control}.\hskip 1em plus 0.5em minus 0.4em\relax IEEE, 2019, pp. 6742--6747.

\bibitem{dean2020sample}
S.~Dean, H.~Mania, N.~Matni, B.~Recht, and S.~Tu, ``On the sample complexity of
  the linear quadratic regulator,'' \emph{Found. Comut. Math.}, vol.~20, no.~4,
  pp. 633--679, 2020.

\bibitem{chen2019system}
Y.~Chen and J.~Anderson, ``System level synthesis with state and input
  constraints,'' in \emph{2019 IEEE 58th Conference on Decision and Control
  (CDC)}.\hskip 1em plus 0.5em minus 0.4em\relax IEEE, 2019, pp. 5258--5263.

\bibitem{lian2021system}
Y.~Lian and C.~N. Jones, ``From system level synthesis to robust closed-loop
  data-enabled predictive control,'' \emph{arXiv, arXiv:2102.06553}, 2021.

\bibitem{chen2020robust}
S.~Chen, H.~Wang, M.~Morari, V.~M. Preciado, and N.~Matni, ``Robust closed-loop
  model predictive control via system level synthesis,'' in \emph{Proc. IEEE
  Conf. Decis. Control}.\hskip 1em plus 0.5em minus 0.4em\relax IEEE, 2020, pp.
  2152--2159.

\bibitem{matni2020robust}
N.~Matni and A.~A. Sarma, ``Robust performance guarantees for system level
  synthesis,'' in \emph{Proc. Am. Control Conf.}\hskip 1em plus 0.5em minus
  0.4em\relax IEEE, 2020, pp. 779--786.

\bibitem{wonham1967optimal}
W.~Wonham, ``Optimal stationary control of a linear system with state-dependent
  noise,'' \emph{SIAM J. Control.}, vol.~5, no.~3, pp. 486--500, 1967.

\bibitem{bujarbaruah2020robust}
M.~Bujarbaruah, U.~Rosolia, Y.~R. St{\"u}rz, X.~Zhang, and F.~Borrelli,
  ``Robust mpc for linear systems with parametric and additive uncertainty: A
  novel constraint tightening approach,'' \emph{arXiv, arXiv:2007.00930}, 2020.

\bibitem{li2021frontiers}
J.~S. Li, C.~A. Alonso, and J.~C. Doyle, ``Frontiers in scalable distributed
  control: Sls, mpc, and beyond,'' in \emph{Proc. Am. Control Conf.}\hskip 1em
  plus 0.5em minus 0.4em\relax IEEE, 2021, pp. 2720--2725.

\bibitem{sieber2021system}
J.~Sieber, S.~Bennani, and M.~N. Zeilinger, ``A system level approach to
  tube-based model predictive control,'' \emph{{IEEE} Contr. Syst. Lett.},
  2021.

\bibitem{chen2021system}
S.~Chen, N.~Matni, M.~Morari, and V.~M. Preciado, ``System level
  synthesis-based robust model predictive control through convex inner
  approximation,'' \emph{arXiv, arXiv:2111.05509}, 2021.

\bibitem{tseng2020system}
S.-H. Tseng, C.~A. Alonso, and S.~Han, ``System level synthesis via dynamic
  programming,'' in \emph{Proc. IEEE Conf. Decis. Control}.\hskip 1em plus
  0.5em minus 0.4em\relax IEEE, 2020, pp. 1718--1725.

\bibitem{prekopa2013stochastic}
A.~Pr{\'e}kopa, \emph{Stochastic programming}.\hskip 1em plus 0.5em minus
  0.4em\relax Springer Science \& Business Media, 2013, vol. 324.

\bibitem{calafiore2011research}
G.~C. Calafiore, F.~Dabbene, and R.~Tempo, ``Research on probabilistic methods
  for control system design,'' \emph{Automatica}, vol.~47, no.~7, pp.
  1279--1293, 2011.

\bibitem{geng2019data}
X.~Geng and L.~Xie, ``Data-driven decision making with probabilistic guarantees
  (part 1): A schematic overview of chance-constrained optimization,''
  \emph{arXiv, arXiv:1903.10621}, 2019.

\bibitem{van2015distributionally}
B.~P. Van~Parys, D.~Kuhn, P.~J. Goulart, and M.~Morari, ``Distributionally
  robust control of constrained stochastic systems,'' \emph{{IEEE} Trans.
  Automat. Contr.}, vol.~61, no.~2, pp. 430--442, 2015.

\bibitem{coulson2021distributionally}
J.~Coulson, J.~Lygeros, and F.~Dorfler, ``Distributionally robust chance
  constrained data-enabled predictive control,'' \emph{{IEEE} Trans. Automat.
  Contr.}, 2021.

\bibitem{stojanovic2003stochastic}
S.~Stojanovic, ``Stochastic finance: An introduction in discrete time,'' 2003.

\bibitem{schildbach2013linear}
G.~Schildbach, P.~Goulart, and M.~Morari, ``The linear quadratic regulator with
  chance constraints,'' in \emph{2013 Eur. Control Conf. ECC 2013}.\hskip 1em
  plus 0.5em minus 0.4em\relax IEEE, 2013, pp. 2746--2751.

\bibitem{Lemarchal2006SBL}
C.~Lemar{\'e}chal, ``S. boyd, l. vandenberghe, convex optimization, cambridge
  university press, 2004 hardback, isbn 0 521 83378 7,'' \emph{Eur. J. Oper.
  Res.}, vol. 170, pp. 326--327, 2006.

\bibitem{NemirovskiArkadi2006ConvexAO}
NemirovskiArkadi and ShapiroAlexander, ``Convex approximations of chance
  constrained programs,'' \emph{SIAM J. Optim.}, 2006.

\bibitem{campi2009scenario}
M.~C. Campi, S.~Garatti, and M.~Prandini, ``The scenario approach for systems
  and control design,'' \emph{Annu. Rev. Control}, vol.~33, no.~2, pp.
  149--157, 2009.

\bibitem{Camp2021TheSA}
M.~T.~C. Camp{\'i}, A.~Car{\`e}, and S.~Garatti, ``The scenario approach: A
  tool at the service of data-driven decision making,'' \emph{Annu. Rev.
  Control.}, vol.~52, pp. 1--17, 2021.

\bibitem{Nemirovski2006ScenarioAO}
A.~Nemirovski and A.~Shapiro, ``Scenario approximations of chance
  constraints,'' 2006.

\bibitem{Calafiore2006TheSA}
G.~C. Calafiore and M.~Campi, ``The scenario approach to robust control
  design,'' \emph{{IEEE} Trans. Automat. Contr.}, vol.~51, pp. 742--753, 2006.

\bibitem{Krmer1995ProbabilityM}
W.~Kr{\"a}mer, ``Probability \& measure: Patrick billingsley (1995): (3rd ed.).
  new york : Wiley, isbn 0-471-0071-02, pp 593,'' \emph{Comput. Stat. Data.
  Anal.}, vol.~20, 1995.

\bibitem{systemID1}
Z.~Mhammedi, D.~J. Foster, M.~Simchowitz, D.~Misra, W.~Sun, A.~Krishnamurthy,
  A.~Rakhlin, and J.~Langford, ``Learning the linear quadratic regulator from
  nonlinear observations,'' in \emph{Thirty-fourth Conference on Neural
  Information Processing Systems (NeurIPS) 2020}, December 2020.

\bibitem{systemID2}
C.~{De Persis} and P.~Tesi, ``Low-complexity learning of linear quadratic
  regulators from noisy data,'' \emph{Automatica}, vol. 128, p. 109548, 2021.

\bibitem{systemID3}
\BIBentryALTinterwordspacing
H.~Mania, ``The sample complexity of simple reinforcement learning,'' Ph.D.
  dissertation, EECS Department, University of California, Berkeley, Aug 2020.
  [Online]. Available:
  \url{http://www2.eecs.berkeley.edu/Pubs/TechRpts/2020/EECS-2020-150.html}
\BIBentrySTDinterwordspacing

\bibitem{systemID4}
S.~Lale, K.~Azizzadenesheli, B.~Hassibi, and A.~Anandkumar, ``Adaptive control
  and regret minimization in linear quadratic gaussian ({LQG}) setting,''
  \emph{arXiv:2003.05999}, 2020.

\bibitem{systemID5}
\BIBentryALTinterwordspacing
Y.~Abbasi-Yadkori and C.~Szepesv\'ari, ``Regret bounds for the adaptive control
  of linear quadratic systems,'' in \emph{Proceedings of the 24th Annual
  Conference on Learning Theory}, ser. Proceedings of Machine Learning
  Research, S.~M. Kakade and U.~von Luxburg, Eds., vol.~19.\hskip 1em plus
  0.5em minus 0.4em\relax Budapest, Hungary: PMLR, 09--11 Jun 2011, pp. 1--26.
  [Online]. Available:
  \url{http://proceedings.mlr.press/v19/abbasi-yadkori11a.html}
\BIBentrySTDinterwordspacing

\bibitem{systemID6}
B.~Cui, Y.~Chow, and M.~Ghavamzadeh, ``Control-aware representations for
  model-based reinforcement learning,'' \emph{arXiv:2006.13408}, 2020.

\bibitem{systemID7}
S.~Arora, E.~Hazan, H.~Lee, K.~Singh, C.~Zhang, and Y.~Zhang, ``Towards
  provable control for unknown linear dynamical systems,'' 2018.

\bibitem{systemID8}
E.~Hazan, H.~Lee, K.~Singh, C.~Zhang, and Y.~Zhang, ``Spectral filtering for
  general linear dynamical systems,'' \emph{arXiv:1802.03981}, 2018.

\bibitem{systemID9}
S.~{Oymak} and N.~{Ozay}, ``Non-asymptotic identification of {LTI} systems from
  a single trajectory,'' in \emph{2019 American Control Conference (ACC)},
  2019, pp. 5655--5661.

\bibitem{systemID10}
S.~Talebi, S.~Alemzadeh, N.~Rahimi, and M.~Mesbahi, ``Online regulation of
  unstable {LTI} systems from a single trajectory,'' \emph{arXiv:2006.00125},
  2020.

\bibitem{systemID11}
C.~Knuth, G.~Chou, N.~Ozay, and D.~Berenson, ``Planning with learned dynamics:
  Probabilistic guarantees on safety and reachability via lipschitz
  constants,'' \emph{arXiv:2010.08993}, 2021.

\bibitem{KOZIN196995}
F.~Kozin, ``A survey of stability of stochastic systems,'' \emph{Automatica},
  vol.~5, no.~1, pp. 95--112, 1969.

\bibitem{confbound}
P.~Auer, ``Using confidence bounds for exploitation-exploration trade-offs,''
  \emph{Journal of Machine Learning Research}, vol.~3, pp. 397--422, 2002.

\bibitem{Chen2007ANG}
X.~Chen, ``A new generalization of chebyshev inequality for random vectors,''
  \emph{arXiv, arXiv:0707.0805}, 2007.

\bibitem{calafiore2014optimization}
G.~C. Calafiore and L.~El~Ghaoui, \emph{Optimization models}.\hskip 1em plus
  0.5em minus 0.4em\relax Cambridge university press, 2014.

\bibitem{Campi_Garatti_2018}
M.~C. Campi and S.~Garatti, \emph{Introduction to the Scenario Approach}.\hskip
  1em plus 0.5em minus 0.4em\relax Society for Industrial and Applied
  Mathematics, Nov 2018.

\bibitem{Calafiore2011ResearchOP}
G.~C. Calafiore, F.~Dabbene, and R.~Tempo, ``Research on probabilistic methods
  for control system design,'' \emph{Automatica}, vol.~47, pp. 1279--1293,
  2011.

\bibitem{Calafiore2017RepetitiveSD}
G.~C. Calafiore, ``Repetitive scenario design,'' \emph{{IEEE} Trans. Automat.
  Contr.}, vol.~62, pp. 1125--1137, 2017.

\bibitem{Calafiore2009OnTE}
------, ``On the expected probability of constraint violation in sampled convex
  programs,'' \emph{J. Optim. Theory. Appl.}, vol. 143, pp. 405--412, 2009.

\bibitem{Calafiore2010RandomCP}
------, ``Random convex programs,'' \emph{SIAM J. Optim.}, vol.~20, pp.
  3427--3464, 2010.

\end{thebibliography}

\end{document}